\documentclass[twocolumn]{aastex631}

\usepackage[colorinlistoftodos]{todonotes}
\usepackage{amssymb}
\usepackage{comment}
\usepackage{multirow,acronym}
\usepackage{natbib}
\usepackage{makecell} 
\usepackage{booktabs}
\usepackage{amsmath}
\usepackage{subfigure}
\usepackage{color}
\usepackage{xcolor}
\usepackage{hyperref}
\usepackage[T1]{fontenc}
\usepackage{graphicx}
\usepackage{bm}
\usepackage{CJK}

\newcommand{\w}{{\rm w}}

\acrodef{OC}{open cluster}
\acrodef{BH}{black hole}
\acrodef{DF}{dynamic friction}
\acrodef{NDF}{negative dynamic friction}
\acrodef{IMF}{initial mass function}
\acrodef{AGN}{active galactic nucleus}

\hypersetup{colorlinks=true, citecolor=blue, 
linkcolor=cyan, urlcolor=magenta}

\usepackage{float}
\usepackage{amsmath,amssymb}
\usepackage{hyperref} % For internal hyperlinks
\usepackage{graphicx} % For included graphics
\usepackage{color}
\usepackage{verbatim}    
\usepackage{enumitem}
\usepackage{natbib}
\usepackage{multirow}
\newcommand{\proptosim}{\mathrel{\vcenter{
 \offinterlineskip\halign{\hfil$##$\cr
 \propto\cr\noalign{\kern2pt}\sim\cr\noalign{\kern-2pt}}}}}
\renewcommand{\b}[1]{\boldsymbol{#1}}

\renewcommand{\min}{\mathrm{min}}
\renewcommand{\max}{\mathrm{max}}
\newcommand{\m}{\mathrm{m}}

\newcommand{\g}{\mathrm{g}}
\newcommand{\G}{\mathrm{G}}
\newcommand{\K}{\mathrm{K}}

\newcommand{\s}{\mathrm{s}}

\renewcommand{\d}{\mathrm{d}}
\renewcommand{\m}{\mathrm{m}}
\newcommand{\f}{\mathrm{f}}
\newcommand{\e}{\mathrm{e}}

\renewcommand{\roman}[1]{\textsc{\MakeLowercase{#1}}}

\newcommand{\p}{{\rm p}}

\newcommand{\figdir}{Figure_petar}

\begin{document}

\title{The effects of star-gas interactions on binary
evolution in open clusters}

\begin{CJK*}{UTF8}{gbsn}
\author[0009-0006-7551-6433]{Jixuan Yang(杨季轩)}
\affiliation{Department of Astronomy, Tsinghua University,
Beijing 100084, China}
%\email{yang-jx24@mails.tsinghua.edu.cn}

\author[0000-0002-6540-7042]{Lile Wang(王力乐)}
\affiliation{The Kavli Institute for Astronomy and Astrophysics, Peking
University, Beijing 100871, China} \affiliation{Department
of Astronomy, School of Physics, Peking University, Beijing
100871, China} %\email{lilew@pku.edu.cn}

\author[0000-0003-0750-3543]{Xinyu Li(李昕宇)}
\affiliation{Department of Astronomy, Tsinghua University,
Beijing 100084, China} %\email{xinyuli@tsinghua.edu.cn}

\author[0000-0001-9037-6180]{Meng Sun(孙萌)} \affiliation{National
Astronomical Observatories, Chinese Academy of Sciences, 20A
Datun Road, Chaoyang District, Beijing 100101, China}
\affiliation{Center for Interdisciplinary Exploration and
Research in Astrophysics (CIERA), Northwestern University,
1800 Sherman Ave, Evanston, IL 60201, USA}
%\email{sunmeng@nao.cas.cn}

\author[0000-0001-8713-0366]{Long Wang(王龙)}
%\email{wanglong8@sysu.edu.cn}
\affiliation{School of Physics and Astronomy, Sun Yat-sen University, Zhuhai 519082, China}

\author[0000-0003-2264-7203]{Rainer Spurzem}
\affiliation{National Astronomical Observatories, Chinese
Academy of Sciences, 20A Datun Road, Chaoyang District,
Beijing 100101, China} \affiliation{Astronomisches
Rechen-Institut, Zentrum für Astronomie, University of
Heidelberg, Möchhofstrasse 12-14, Heidelberg D-69120,
Germany} \affiliation{The Kavli Institute for Astronomy and
Astrophysics, Peking University, Beijing 100871, China}
%\email{spurzem@ari.uni-heidelberg.de, spurzem@nao.cas.cn}

\correspondingauthor{Xinyu Li} \email{xinyuli@tsinghua.edu.cn}

\correspondingauthor{Meng Sun} \email{sunmeng@nao.cas.cn}
\begin{abstract} Star-gas interactions can provide gravitational feedback that influences the dynamical evolution of stellar clusters, through processes such as dynamical friction (DF) and its 
%non-dissipative 
counterpart, negative dynamical friction (NDF). Using the \texttt{PeTar} code, we perform direct $N$-body simulations of an open cluster initially containing $10^4$ stars, evolving
within a gaseous medium spanning a range of ambient densities. Our results
demonstrate that NDF associated with stellar outflows interacting with the surrounding gas can enhance the rate of cluster expansion, preferentially transporting stars toward the cluster outskirts. This behavior is accompanied by a more rapid decline in the number of binaries composed of a neutron star and a main-sequence star.
%significantly accelerates cluster evaporation by expelling stars to the cluster outskirts, forming an expanding shell. This process leads to a faster decrease in the number of binary systems containing a compact object and a main-sequence star. NDF更容易让有强外流的compact object（比如NS）相关双星减少，但是MS-WD数量W由恒星演化生成主导，所以统称主序星+致密天体binaries下降，不准确
% Conversely, DF
% drives mass segregation and can increase the binary
% population by enhancing stellar encounters in the cluster
% core. 
A statistical analysis of binary orbital parameters further indicates that, compared to DF-dominated evolution, NDF tends to retain systems with larger semi-major axes and lower eccentricities. Outflow-ambient gas interactions can modify the dynamical processing of binaries in star clusters, leading to changes in the survival fraction and composition of the remaining binary population.

%Outflow-ambient gas interactions generally shape the binary population in clusters, and eventually impact the predictions on gravitational wave source populations that are related to clustered stars.
%我不建议在abstract提到引力波内容，因为我们没做GW source population modeling
\end{abstract}


\section{Introduction}
\label{sec:intro}

%\lilewnote{I thoroughly re-wrote your introduction section,
%which used to be quite far away from being
%publishable. Please be noted that the introduction section
%should state the purpose of the study by identifying (1) why
%the topic is important, (2) why we should do it (i.e., what
%are the problems and defects of previous works), and (3)
%very briefly, the intended improvements of our work. You can
%first write a few sentences for each paragraph for the idea,
%put it into the prompt of a GenAI, and polish the AI output
%yourself.}

Binaries play a fundamental role for the dynamical and energetic evolution of star cluster, because hard binaries interact with cluster field stars in superelastic few-body interactions, and soft binaries could act as cooling through their ionization \citep{1991ApJ...370..567G, 1992PASP..104..981H, 2003MNRAS.343..781G, 2021MNRAS.508.4385A}.
%Binary star systems play a fundamental role in stellar dynamics, serving as crucial laboratories for understanding multiple astrophysical processes \citep{Tauris2023,Marchant2024,2025CoSka..55c..21B}. 
Their evolution drives key
mechanisms including mass transfer episodes \citep{Soberman1997}, common-envelope
phases \citep{Iben1993,Taam2006,2013A&ARv..21...59I}, Type Ia supernova
progenitors \citep{Hillebrandt2000,Wang2012,2014ARA&A..52..107M,Livio2018,2023RAA....23h2001L}, 
and the formation of exotic objects like blue stragglers
\citep[e.g.][]{1986AJ.....92.1364M,Bailyn1995,Sills2010,2025ARA&A..63..467M}. 
Furthermore, binary
interactions significantly influence planet formation in
dense environments \citep{2007A&A...462..345D,Duchene2013,Kraus2016} and represent
the primary progenitors for gravitational wave (GW) events
detected by advanced facilities \citep{Postnov2014,Abbott2016,Mapelli2016,Mandel2022a,Mandel2022b,2023PhRvX..13d1039A}.
 
There are over $\sim 10^3$ cataloged open clusters (OCs) in
the Milky Way \citep{2018A&A...618A..93C}, providing a
statistically rich and dynamically diverse laboratory for
studying the evolution of binary populations. Since most stars
form in clustered environments \citep{2003ARA&A..41...57L},
the dynamical processes within OCs can play an important role in
shaping binary populations. The relatively high stellar densities in clusters
enhance close encounters and dynamical interactions
\citep{2019MNRAS.486.5008A}; however, the resulting occurrence rates of GW
events from compact object mergers in clusters remain
debated. While high densities are expected to increase encounter
rates, the stochastic nature of cluster dynamics can also produce
highly eccentric orbits and eject binaries
before merger \citep[see also][] {2014MNRAS.440.2714B,
2016ApJ...831..187A, 2023ApJ...944..112W}. Consistent with this picture, numerical
studies have examined the potential contribution of dynamical
interactions in open clusters to the formation and evolution of
compact binaries relevant for GW sources
\citep{Rodriguez2018,Kumamoto2019,Rastello2019}.

The interaction between OCs and their ambient interstellar
medium (ISM) represents a crucial, yet often overlooked,
aspect of cluster evolution. OCs form from giant molecular
clouds and experience an early embedded phase,
followed by continued interactions with the surrounding gas as they evolve within the Galactic disk \citep{1987ARA&A..25...23S,Kroupa2001,2003ARA&A..41...57L,Goodwin2006}. During these
stages, star-gas interactions can influence the dynamical
evolution of clusters. The theoretical foundation for gas dynamical friction (DF)
was established by \citet{1999ApJ...513..252O}, who showed
that a gravitating object moving through a gaseous medium experiences
a decelerating force due to the formation of an overdense wake. Recent theoretical
advances by \citet{2020MNRAS.492.2755G} and numerical
simulations by \citet{Li_NDF_2020} have revealed
that stars with strong outflows can instead generate negative
dynamical friction (NDF), whereby an underdense cavity
leads to a net acceleration. Such effects are expected to be most
relevant for stars with substantial mass loss, particularly during
post-main-sequence evolution \citep{Smith2014}.

Star-gas interactions occur in active galactic nuclei (AGNs) when a nuclear star cluster interacts with a central gaseous disk. Traditionally, classical dynamical friction and ram-pressure approximations have been employed to model these interactions \citep[e.g.,][]{2012ApJ...758...51J, 2018MNRAS.476.4224P}. More recently, this process has gained renewed attention as a potential channel for gravitational-wave sources \citep{2020ApJ...889...94M}. NDF could significantly change the trapping and transport of massive stars and black holes within AGN disks.

Understanding how DF and NDF affect binary populations in
OCs is crucial for a range of astrophysical contexts. Binary
fractions in OCs show significant variation
\citep[e.g.][]{2011A&A...528A.144K, 2024ApJ...971...71J},
and gas-mediated friction could substantially modify binary
formation and dissolution rates. Previous N-body simulations
by \citet{2025ApJ...978...87L} investigated OC evolution
with gas interactions but were limited by cluster size
($N \sim 1.3\times 10^3$), yielding insufficient binary
statistics for robust analysis. 
Recent observational studies have begun to characterize binary populations in OCs, including their fractions, orbital properties, and evolutionary outcomes, highlighting the
need for larger-scale simulations \citep{Mathieu2004,Geller2009,Geller2015,Geller2012,Leiner2021}.  
Factors influencing the binary population in clusters and their orbital parameters include cluster-driven
dynamical processes, tidal stripping, and gas dynamical effects \citep{1995ApJ...445L.133R, 1996MNRAS.282.1064H,Fregeau2009,Leigh2016,Fragione2020,Kremer2020,Ye2020,Gonzalez2022,Weatherford2023,Atallah2024,Kiroglu2025}. In this work, we focus on the impact of the last channel.

We increase the simulation scale to $N = 10^4$
stars using the \texttt{PeTar} $N$-body code 
\citep{2020MNRAS.497..536W}, with modifications
in external force module and stellar evolution module. 
This larger
particle count allows us to retain a statistically significant binary
population, enabling detailed analysis of how DF and NDF affect binary parameters, survival rates, and orbital properties across a range of ambient gas densities. Our approach
incorporates stellar evolution prescriptions and the
respective accelerations from both DF and NDF, providing a systematic investigation of gas-mediated friction
effects on binary populations in open clusters.

The paper is organized as follows. In Section~ \ref{sec:method} we describe the simulation setup and methodology. Section
\ref{sec:results} presents the stellar dynamical properties. In Section \ref{sec:binary} we focus on the evolution of
binary populations under different friction regimes. Section \ref{sec:summary} summarizes our main conclusions
and discusses their implications.

\section{METHOD}
\label{sec:method}

\subsection{N-body Code}
\label{sec:method.1}
We perform numerical simulations using the \texttt{PeTar} N-body integrator \citep{2020MNRAS.497..536W} to track the detailed phase-space evolution of each star over 200 Myr during the cluster evolution. The simulations
are carried out in a large computational domain
($1000~\mathrm{pc}$ on a side), chosen to avoid boundary effects during the cluster evolution.

% NOTE: We mention a large computational box (1000 pc) to avoid boundary effects. Please let me know if PeTar uses any periodic or reflective boundary conditions, or if this box size is purely bookkeeping. we can adjust the wording here if needed.

%We perform our simulations using the \texttt{PeTar} N-body
%integrator \citep{2020MNRAS.497..536W}, which calculates the
%gravitational interactions between particles while
%supporting user-defined additional physics modules. For this
%work, we have implemented both dynamical friction
%prescriptions and comprehensive stellar evolution models
%within the \texttt{PeTar} framework. 

% Unlike the IAS15
% integrator employed by \citet{2025ApJ...978...87L}, we adopt
% a 4th-order leap-frog scheme. This symplectic integrator
% provides superior energy conservation properties compared to
% IAS15, resulting in significantly reduced displacement of
% the open cluster's centroid. While sacrificing some accuracy
% in individual particle trajectories, the leap-frog method
% substantially improves computational efficiency, enabling
% our larger-scale simulations. We observe centroid
% displacements on the order of $100~\rm{pc}$ when using
% IAS15, compared to approximately $10~\rm{pc}$ with the
% leap-frog scheme. Given our $500~\rm{pc}$ simulation
% bounding box, this level of error is acceptable. The cluster
% center is determined using the median position along each
% axis to minimize the influence of outliers.
%The boundaries of simulations are given by a boxes with side 
%lengths of $1000~\rm{pc}$.

\begin{figure}
    \centering
    \includegraphics[width=1\linewidth]{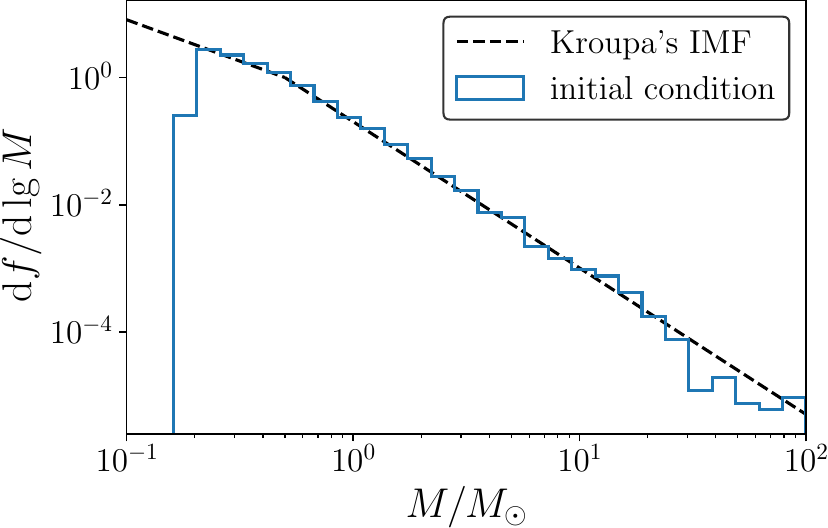}
    \caption{Mass distribution of the initial stellar population in this work. The blue step line shows the sampled initial stellar mass distribution, while the black dashed line indicates the \citet{2002Sci...295...82K} IMF (Eq.~\eqref{eq:IMF_kroupa}), shown in units of $\mathrm{d}f/\mathrm{d} \lg M$.}
    \label{fig:IMF}
\end{figure}

Our initial conditions are generated using \texttt{MCLUSTER}
\citep{mcluster}, creating an open cluster with
10,000 stars. 
The stellar mass distribution follows the
broken power-law initial mass function of 
\citet{2002Sci...295...82K}, sampled
over the mass range $0.2$--$100~M_{\odot}$, as shown in Figure~\ref{fig:IMF}. The lower-mass cutoff is adopted to
reduce computational cost while preserving the global
dynamical properties of the cluster: 
\begin{equation}\label{eq:IMF_kroupa}
    \xi(M)\propto \left\{
    \begin{array}{ll}
    M^{-1.3}, ~M<0.5~M_{\odot},\\
    M^{-2.3},~M\geq 0.5~M_{\odot}.
    \end{array}
    \right.
\end{equation}
The cluster is initialized with a half-mass radius of
$3~\mathrm{pc}$ and follows a King density profile of
\citet{1966AJ.....71...64K} with a concentration parameter
$W_0=5$, typical of young OCs. 
There are 2,000 binaries in the initial cluster, corresponding to an initial binary fraction of 40\%
%The cluster also initially contains 2,000 binaries 
\citep{2010MNRAS.401..577S, 2023A&A...675A..89D}.

We adopt standard initial binary parameter distributions commonly used in open-cluster simulations: orbital periods are drawn from \citet{2008LNP...760..181K} period distribution, which are then converted to semi-major axes, eccentricities follow a thermal distribution, and mass ratios are assigned using a mass-dependent pairing scheme. The overall binary fraction and pairing prescription follow \citet{2012Sci...337..444S}, with higher binary fractions for massive stars.

\subsection{Dynamical Friction and Negative Dynamical Friction}
\label{sec:method-fric}
We implemented both DF and NDF prescriptions within the \texttt{PeTar} framework. Stars without significant outflows are assumed to experience the conventional
dynamical friction given by
\citet{1999ApJ...513..252O}:
\begin{equation}\label{eq:a_DF}
    a_{\rm{DF}}=\dfrac{4\pi G^{2}\rho_0 M}{c_s^2}I_{\rm DF}(\mathcal{M}),
\end{equation}
where the acceleration is applied antiparallel to the stellar velocity relative to the gas. The dimensionless function $I_{\rm DF}(\mathcal{M})$ is given by
\begin{equation}
I_{\rm DF}(\mathcal{M})=\dfrac{1}{\mathcal{M}^{2}}\times\left\{
\begin{array}{ll}
{{\ln\left[\Lambda\left(1-\dfrac{1}{\mathcal{M}^{2}}\right)^{1/2}\right]\,,\
  \mathcal{M}>1\ ;}}\\ 
{\dfrac{1}{2}\ln\left(\dfrac{1+\mathcal{M}}{1-\mathcal{M}}\right)-\mathcal{M}\
  ,\ \mathcal{M}<1\ .} 
\end{array}\right.
\label{eq:I_DF}
\end{equation} 
Here $\rho_0$ is the gas density, and $\Lambda \equiv
b_{\max}/b_{\min}$ is the Coulomb factor with
$b_{\max}=100~\mathrm{pc}$ (typical molecular cloud size)
and $b_{\min}$ is given by the Bondi radius, $r_{\rm B}=2GM/c_s^2$. 
The Mach number $\mathcal{M}\equiv
v_*/c_s$ uses the sound speed $c_s$ of the ambient medium, 
and $v_*$ represents the stellar velocity relative to the
ambient gas.

For stars with significant outflows, we implement a
prescription for NDF following \citet{2020MNRAS.492.2755G} and \citet{Li_NDF_2020}: 
\begin{equation}
\begin{aligned} 
a_{\mathrm{NDF}}&\simeq\pi
                  G\rho_{0}\int_{0}^{\pi}\mathrm{d}\theta\,\cos\theta\,\sin\theta\,R_{s}
  \\ 
&\times\left\{\dfrac{3}{2}\left[1+\dfrac{2u(1-\cos\theta)}{R_{s}^{2}\sin^{2}\theta/R_{0}^{2}}\right]^{2}-2\left[1+\dfrac{u^{2}}{R_{s}^{2}/R_{0}^{2}}\right]\right\}.
\end{aligned}
\end{equation} 
The standoff distance
$R_{0} =
[\dot{m}_{w}v_{w}/(4\pi\rho_{0}v_{\ast}^{2})]^{1/2}$ depends
on the mass-loss rate $\dot{m}_{w}$ and wind velocity
$v_{w}$. The radius of contact discontinuity is
$R_s\simeq R_0
[3(1-\theta\,\mathrm{cot}\,\theta)/\sin^{2}\theta]^{1/2}$
and velocity ratio is $u\equiv v_*/v_{w}$. For the low
velocity ratios typical in open clusters ($u\ll 1$), this
simplifies to $a_{\mathrm{NDF}}\simeq 8.18G\rho_0 R_0$
\citep{Li_NDF_2020}.
\begin{equation}\label{eq:a_NDF}
\begin{aligned} 
    a_{\rm{NDF}}&=8.18G\rho_0 \sqrt{\dfrac{\dot{m}_w v_w}{4\pi\rho_0 v_*^2}}\\
    &=8.18G\sqrt{\dfrac{\dot{m}_w v_w \rho_0}
    {4\pi}}\dfrac{1}{c_s}I_{\rm NDF}(\mathcal{M}),
\end{aligned}
\end{equation}
\begin{equation}
    I_{\rm NDF}(\mathcal{M})=\dfrac{1}{\mathcal{M}}.
\label{eq:I_NDF}
\end{equation}
The resulting acceleration is applied parallel to the stellar velocity relative to the gas.

At sufficiently high stellar velocities, NDF can be suppressed.
The critical point is determined by
comparing the ambient gas ram pressure $p_0\sim \rho_0 v_*^2$
with the stellar wind pressure evaluated at the sonic point:
\begin{equation}
p_s=\left(1+\frac{1}{\gamma}\right)\frac{\dot{m}_w c_{s,w}^5}{\pi G^2 M^2},
\end{equation}
where $\gamma =5/3$ and the wind sound speed is defined as
$c_{s,w}= \sqrt{2k_BT_w/m_p}$, with $T_w$ being the wind temperature. When $p_0\leq p_s$, stellar winds are able to establish a bow shock, leading to the formation of an underdense cavity and the onset of NDF; when $p_0>p_s$, wind is choked by the ambient medium and the NDF contribution is suppressed, such that $I_{\rm{NDF}}$ 
drops to zero. This criterion defines the transition between the active and suppressed NDF regimes in our simulations.

%As the timestep of the \texttt{PeTar} integrator is determined by $a$, $\dot{a}$, $\ddot{a}$ and $\dddot{a}$, we design Laurent polynomials of order 5 to -2 to deal with the jump of $I_{\rm{DF}}$ between $\mathcal{M}< 1$ and $\mathcal{M}> 1$, and also Laurent polynomials of order 6 to -1 for the jump of $I_{\rm{NDF}}$ between $p_0< p_s$ and $p_0> p_s$. And we also use 4th polynomials to replace original $I_{\rm DF}$ and $I_{\rm NDF}$ (Eq. \eqref{eq:I_DF} \& \eqref{eq:I_NDF}) when $\mathcal{M}\to 0$ to improve the computing accuracy. All these polynomials satisfy that $a$, $\dot{a}$, $\ddot{a}$ and $\dddot{a}$ are continuous.

\begin{figure}
    \centering
    \includegraphics[width=1\linewidth]{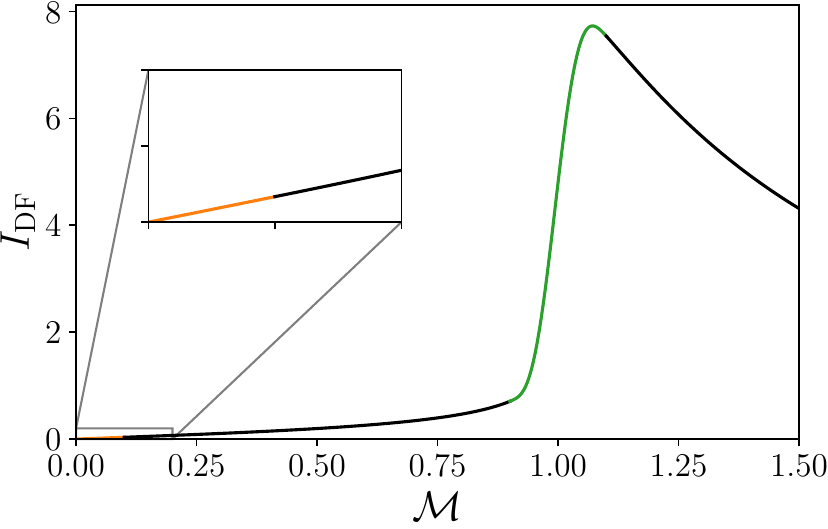}
    \caption{$I_{\rm DF}(\mathcal{M})$ for $\ln{\Lambda}=10$. 
    The black curve shows the analytic expression for $I_{\rm DF}$ given by Equation~\eqref{eq:I_DF},
    while the orange and green curves show the polynomial approximations adopted in our simulations to ensure numerical smoothness.}
    \label{fig:a_DF}
\end{figure}

\begin{figure}
    \centering
    \includegraphics[width=1\linewidth]{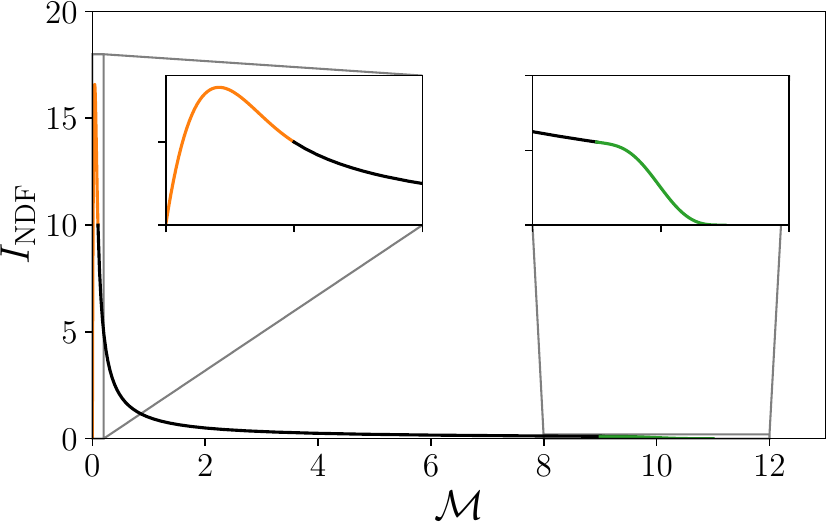}
    \caption{$I_{\rm NDF}(\mathcal{M})$ for $\mathcal{M}_{\rm{crit}}=10$, where $\mathcal{M}_{\rm crit}$ is the Mach number at which $p_0=p_s$.
    The line styles and colors follow those in Figure~\ref{fig:a_DF}, with the black curve corresponding to the analytic expression and the colored curves showing the polynomial approximations.}
    \label{fig:a_NDF}
\end{figure}

Because the timestep of the \texttt{PeTar} integrator depends on the acceleration and its time derivatives up to third order, numerical discontinuities in the force prescription can lead to integration instability. To ensure numerical smoothness across transition regimes, we construct smooth
polynomial interpolations for both the DF and NDF kernels. Specifically, we smooth the transition in $I_{\rm DF}$ across $\mathcal{M}=1$ and the transition in $I_{\rm NDF}$ across $p_0 = p_s$ using tailored polynomial interpolations (including finite negative-order terms where appropriate). In addition, to avoid numerical singularities as $\mathcal{M}\to 0$, we replace the original expressions for $I_{\rm DF}$ and $I_{\rm NDF}$ (Equations~\ref{eq:I_DF} and \ref{eq:I_NDF}) with fourth-order polynomial expansions in this limit. All interpolations are constructed such that the acceleration and its first three time derivatives are continuous. The resulting interpolated forms of $I_{\rm DF}$ and $I_{\rm NDF}$ are shown in Figures~\ref{fig:a_DF} and \ref{fig:a_NDF}, respectively.

%Our interpolated $I_{\rm{DF}}$ and $I_{\rm{NDF}}$ are shown as Figure \ref{fig:a_DF} and \ref{fig:a_NDF} respectively.

%The flow field establishment occurs on Bondi timescales,
The characteristic timescale of the gas flow is the Bondi timescale,
which is much shorter than the the typical dynamical timescales associated with open cluster evolution. We therefore treat DF/NDF accelerations as instantaneous in our simulations. Based on this assumption, we simulate three interaction modes: ``NONE'' (no gas interactions),
``DF'' (including only dynamical friction), and ``NDF'' (including NDF
for outflowing stars and DF for stars without significant outflows).

\begin{deluxetable*}{ccccc}
    \centering
    \caption{Summary of simulation parameters}
    \label{tab:sims_list}
    
        \tablehead{
        \colhead{ID} & \colhead{\makecell{$\rho_0^{\dagger}$\\($m_p~\rm cm^{-3}$)}} & \multicolumn{2}{c}{Models$^{\ddagger}$} & \colhead{Description} \\
        \cmidrule(lr){3-4}
         &  & DF & NDF &  }
        \startdata
        0 & 0 & \multicolumn{2}{c}{NONE} & No interaction between stars and gas. \\
        1 & 30 & DF-D3E1$^{\S}$ & NDF-D3E1 & CNM in the Milky Way. \\
        2 & 300 & DF-D3E2 & NDF-D3E2 & Dense neutral HI clouds. \\ 
        3 & $3\times 10^3$ & DF-D3E3 & NDF-D3E3 & Diffuse molecular cloud. \\
        4 & $3\times 10^4$ & $*$ & NDF-D3E4 & Dense molecular cloud cores. \\
        5 & $3\times 10^6$ & $*$ & NDF-D3E6 & Densest molecular cloud cores. \\
        6 & $3\times 10^8$ & $*$ & NDF-D3E8 & 
        AGN-like dense gaseous environment (included for comparison only).\\
        \enddata

    \tablecomments{
    % $^{\dagger}$ At the same ambient gas density, $|a_{\rm{DF}}|$ is 
    % much larger than $|a_{\rm{NDF}}|$. When $\rho_0$ is higher than
    % $3\times10^3 ~m_p/\rm cm^3$, timestep is too small in DF mode. So we focus on
    % the effect of NDF and run 6 simulations from $\rho=30~ m_p/\rm cm^3$ to
    % $3\times 10^8~m_p/\rm cm^3 $.
${\dagger}$: $\rho_0$ is the initial ambient gas density.\\
${\ddagger}$: DF/NDF denotes dynamical friction / negative dynamical friction models; NONE indicates no gas-star interaction.\\
${\S}$: The suffix (e.g., D3E1) denotes specific parameter sets.\\
$*$: Asterisk indicates that DF simulations were not performed at these densities because the required time steps were prohibitively small.
    }
\end{deluxetable*}

We assume the cluster resides in cold neutral medium (CNM)
with $T\simeq 100\ \rm{K},~\gamma=5/3$ and explore gas number densities from
$\rho_0 = 30\ m_p\ \rm{cm^{-3}}$ (Milky Way CNM) to
$\rho_0 = 3\times 10^8\ m_p\ \rm{cm^{-3}}$, representing
extremely dense gaseous environments. Specifically, we consider $\rho_0 = 30$, $300$, $3\times 10^3$, $3\times 10^4$,
$3\times 10^6$, and $3\times 10^8 \ m_p\ \rm{cm^{-3}}$.

\subsection{Stellar Evolution}
\label{sec:method.3}
We employed a simplified, parameterized stellar evolution model designed to capture the dominant evolutionary stages relevant for mass loss and stellar outflows, rather than to model detailed stellar structure or binary mass transfer.
The mergers and spins are also not taken into account.
The model includes the main sequence (MS), red giant branch (RGB), and asymptotic giant branch (AGB) phases, followed by transitions to compact objects.
%Stellar evolution in this work is highly simplified to prevent the influence of gas-star interactions from being masked with by dynamical effects introduced by stellar evolution. 
%Our stellar evolution model incorporates main sequence (MS),
%red giant branch (RGB), and asymptotic giant branch (AGB)
%phases, with transitions to compact objects. 
The MS lifetime is approximated as
$t_{\mathrm{MS}}\simeq10^{10}\;\mathrm{yr}\times(M/M_{\odot})^{-2.5}$.
The RGB duration is parameterized as
\begin{equation}
t_{\mathrm{RGB}}=\left\{
  \begin{array}{ll}
    10^{5}~\mathrm{yr}\ & M >15~M_{\odot} \\
    3\times10^{8}\times(M/M_{\odot})^{-4}~\mathrm{yr}\ ,
                        & M \leq 15~M_{\odot}\ .
  \end{array}\right. 
\end{equation}
These scalings are chosen to reproduce the approximate mass dependence of post-main-sequence evolutionary timescales. 

Stars with $M<8\ M_{\odot}$ undergo AGB phases
($\sim10^5\ \mathrm{yr}$) before becoming white dwarfs,
while more massive stars experience supernovae (SN)
($\sim10^4\ \mathrm{yr}$) forming neutron stars ($8~ M_{\odot}\leq M < 25~ M_{\odot}$) or black
holes ($M\geq 25~M_{\odot}$).

%AGB winds are assumed to have a velocity $v_w=30\ \rm{km\ s^{-1}}$ \citep{AGB_wind_velocity}, while SNe ejecta are launched at $3\times 10^4\ \rm{km\ s^{-1}}$. Although SNe outflows are intrinsically anisotropic, we neglect natal kicks in order to isolate the effects of NDF, applying $a_{\rm NDF}$ along the instantaneous stellar velocity. We further assume that the ambient gas density remains uniform and constant in time. This approximation neglects gas removal by SN feedback and local clearing by binaries, and is adopted to focus on the dynamical impact of DF and NDF under controlled gas conditions.

% 2026.1.2 1:27 AM record

We model stellar mass ejection across different evolutionary phases using a simplified, parameterized prescription, with the goal of providing order-of-magnitude outflow properties relevant for NDF, rather than a detailed treatment of stellar wind physics.

For main-sequence stars, we adopt a simple mass-dependent wind prescription to estimate stellar outflows relevant for NDF. The mass-loss rates follow \citet{Reimers_mass_loss}:
\begin{equation} 
\dot{m}_w\simeq 4\times10^{-13}M_{\odot}\
\mathrm{yr^{-1}}\times
\dfrac{L}{L_{\odot}}\dfrac{R}{R_{\odot}}\dfrac{M_{\odot}}{M}. 
\end{equation}
Using $L\propto M^{3.5}$ \citep{mass-luminosity_relation} and $R\propto M^{0.8}$ \citep{mass-radius_relation} yields
$\dot{m}_w\propto (M/M_{\odot})^{3.3}$. Wind velocities scale as $v_w=400\ \mathrm{km\ s^{-1}}\times(M/M_{\odot})^{0.6}$,
consistent with solar values and observations of massive O stars. The wind temperatures follow \citet{MS_wind_model}:
\begin{equation} 
T_w=\dfrac{2G\overline{\mu}}{\gamma
  k_B}\left(\dfrac{c_s}{v_{\mathrm{esc}}}\right)^2\dfrac{M_*}{R_*}\sim 2.2\times 10^6~{\rm K}\times  \left(\dfrac{M}{M_{\odot}}\right)^{0.2} , 
\end{equation}
with a fiducial ratio $c_s/v_{\mathrm{esc}}\sim 0.4$.

For evolved stars, we adopt fiducial outflow parameters. 
We assume RGB stars have $v_w\approx 30~{\rm km~s^{-1}}$, $\dot{m}_w\approx 3\times 10^{-9}M_{\odot}~{\rm yr^{-1}}$
and $T_w\sim 1\times 10^6~\rm K$ \citep{wind_rgb}.

The mass-loss rates during the AGB and SN phases 
are treated as constant and are determined by the 
difference between the stellar mass at the end of the 
RGB phase and the final remnant mass. Final remnant masses are $\min\{0.5\ M,\ 1.0\,M_{\odot}\}$ for WD progenitors, $\min\{0.1\ M,\ 1.5\ M_{\odot}\}$ for NS progenitors and $5~M_{\odot}$ for BH progenitors. 
AGB winds are assumed to have a characteristic velocity $v_w = 30\,\mathrm{km\,s^{-1}}$ \citep{AGB_wind_velocity}, while SN ejecta is assigned a characteristic velocity $v_w=3\times 10^4~{\rm km~s^{-1}}$.
Although supernova outflows are intrinsically anisotropic, we neglect natal kicks in order to isolate the effects of NDF, applying $a_{\rm NDF}$ along the instantaneous stellar velocity.

For compact objects, neutron stars eject material at $v_w = 3\times10^{4}~\mathrm{km~s^{-1}}$ \citep{wind_velocity_NS}, with the associated $\dot{m}_w\sim3\times 10^{-9} M_{\odot}~{\rm yr^{-1}}$ and $T_w\sim 10^{8.6}~\rm K$ adopted from \citet{mass_loss_NS}.
We neglect WD and BH outflows, as they do not provide significant steady mass loss relevant for NDF. 

The adopted outflow parameters for the relevant evolutionary phases are summarized in Table~\ref{tab:outflow}. We collectively refer to all mass-ejection processes as outflows for simplicity.
This simplified prescription is adopted for computational efficiency and does not affect our qualitative conclusions, which depend primarily on the presence and strength of stellar outflows.

\begin{deluxetable*}{cccc}
  \caption{Adopted outflow parameters in different evolutionary phases. } 
\label{tab:outflow}
\tablehead{
  \colhead{Evolution phase} &
  \colhead{$\dot{m}_w/(M_{\odot}~\mathrm{yr^{-1}})$}&
  \colhead{$v_w/(\mathrm{km~s^{-1}})$} &
  \colhead{$T_w/(10^6~\mathrm{K})$}}
\startdata
MS & $4\times10^{-13}\times\left(M/M_{\odot}\right)^{3.3}$ &
$400\times\left(M/M_{\odot}\right)^{0.6}$ & $2.22\times
(M/M_{\odot})^{0.2}$ \\
RGB & $3\times 10^{-9}$ & $30$ & $1$ \\
AGB & $\dagger$ & 30 & $*$ \\
SN ejecta$^{\ddagger}$ & $\dagger$ & $3\times 10^4$ & $*$ \\
NS & $3\times 10^{-9}$ & $3\times 10^4$ & $10^{2.6}$ \\
\enddata
\tablecomments{
$\ddagger$: For the purpose of computing gas-mediated dynamical effects, SN ejecta are treated as an effective outflow.\\
$\dagger$: The mass-loss rates during the AGB and SN phases 
are treated as constant and are determined by the 
difference between the stellar mass at the end of the 
RGB phase and the final remnant mass.\\
$*$: We assume $p_s \gg p_0$ during the AGB and SN phases. Consequently, we do not explicitly calculate $p_s$ and therefore do not specify the wind temperature in these two phases.
}
\end{deluxetable*}

Finally, we assume that the ambient gas density remains uniform and constant in time. This approximation neglects gas removal by supernova feedback and local clearing by binaries, and is adopted to focus on the dynamical impact of DF and NDF under controlled gas conditions.

\subsection{Binary Evolution and Analysis}
\label{sec:method-bin}

%Binary systems play a crucial role in cluster dynamics and evolution. 
%However, close binaries are difficult to form 
We identify bound stellar pairs by evaluating their total
energy in the center-of-mass frame,
\begin{equation}\label{eq:E_tot}
  E_{\rm tot} = -\dfrac{G m_1 m_2}{|\vec{r}_1 - \vec{r}_2|}
  + \dfrac{m_1 m_2(\vec{v}_1 - \vec{v}_2)^2}{2(m_1 + m_2)}\ .
\end{equation}
Pairs with $E_{\text{tot}} < 0$ are classified as gravitationally bound. Binary candidates are identified using the \texttt{petar.data.process} module. A bound pair is retained as a binary if the two stars are mutual nearest neighbors and their separation is smaller than $5~\rm pc$.
% To optimize computational efficiency, we first
% identify candidate binaries using a KD-tree algorithm to
% find stars that are the closest neighbors to each other with a separation of less than 1 pc, then compute energies
% only for these candidate pairs \citep{2020JOSS....5.3094V}.

For each identified binary, we track the instantaneous orbital parameters, including semi-major axis
$a$, eccentricity $e$ and total system mass $M_{\text{tot}}$
% , mass ratio $q$ (defined as $m_2/m_1 \leq 1$), and orbital inclination
. These quantities are computed from the relative
positions and velocities of the two components at each output time.

Gas--star interactions are incorporated by applying the DF and NDF acceleration to each component star
independently. This treatment assumes that the presence of a binary companion does not strongly
modify the local gas response. This approximation is expected to
be valid for wide binaries, but may break down for close systems
where the binary orbit can perturb the surrounding gas distribution \citep{2022ApJ...932..108W}.

We classify binaries according to their stellar composition: MS--MS, MS--WD, MS--NS,
and MS--BH. This classification enables a
direct comparison of how different stellar types and their associated outflow properties influence binary survival and orbital evolution under gas-mediated interactions.
All stars are initialized as zero-age MS objects at the beginning of the simulation. Over the adopted evolutionary timescale of $200~\rm Myr$,  the majority of stars remain on the MS, and MS--MS binaries therefore dominate the overall binary population. We place particular emphasis on MS--NS binaries. NS experience strong NDF due to their powerful
outflows, making these systems especially sensitive to NDF-driven orbital evolution and disruption. This population is also directly relevant for gravitational wave source demographics \citep{2021PhRvX..11b1053A}.
%Factors influencing the binary population in clusters and their orbit parameters include three-body interactions, tidal stripping, and gas dynamical effects \citep{1995ApJ...445L.133R, 1996MNRAS.282.1064H}. In this work, we investigate the impact of the latter by varying the ambient gas density. 

In general, NDF acts to increase
binary orbital energy, preferentially disrupting wider systems,
whereas DF removes orbital
energy and tends to harden bound pairs. We quantify these competing effects statistically by comparing binary populations across a range of ambient gas densities.

\section{Cluster evolution and stellar dynamics}
\label{sec:results}

%Our simulations using the \texttt{PeTar} integrator track the detailed phase-space evolution of each star over a 200 Myr period, providing comprehensive data on stellar positions, velocities, and evolutionary states. Given that massive stars ultimately evolve into compact objects, and that among these only neutron stars (NSs) maintain substantial outflows within our model framework, we focus on the dynamical behavior of NSs.

\subsection{Neutron star expulsion via NDF}

We first examine how NDF alters the orbital evolution of NSs in embedded star clusters, and whether it can efficiently remove compact remnants from the cluster independently of supernova kicks. To isolate the role of NDF, we compare simulations with DF-only, NDF, and gas-free (NONE) prescriptions across a range of ambient gas densities.

\begin{figure*} \centering
\includegraphics[width=1\linewidth]{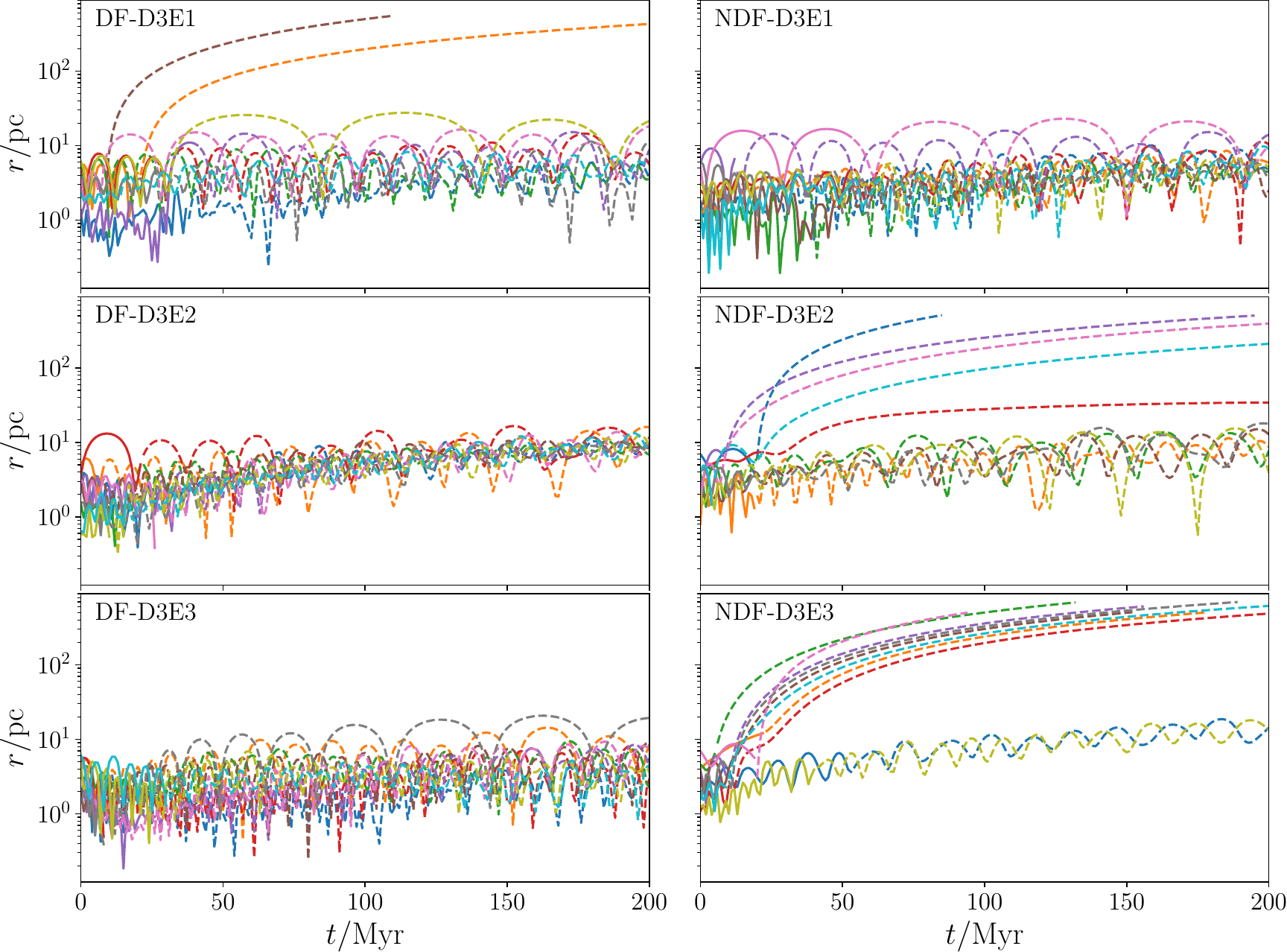}
%    \caption{The track of stars evolving from MS to NS, plotted as a function of time on the horizontal axis and the distance from the cluster center on the vertical axis. Due to the large number of stars, only 10 randomly selected trajectories are shown for each simulation. The breaks in the curve indicate stars leaving the bounding box of the simulation, evaporating from the cluster. Solid lines denote stars in the main sequence phase, while dashed lines indicate neutron stars. Different rows show the results in different density from $30 ~m_p~\rm{cm^{-3}}$ to $3\times 10^3~m_p~\rm{cm^{-3}}$. Left and right columns compare the difference between DF mode and NDF mode respectively.}
\caption{Radial trajectories of stars that transition from the MS phase to the NS phase, shown as a function of time (horizontal axis) and distance from the cluster center (vertical axis). Solid lines represent stellar trajectories during the MS phase, while dashed lines indicate the subsequent trajectories after the stars have become NSs. Only 10 randomly selected trajectories are shown for each simulation for clarity. Breaks in the curves mark objects that
leave the simulation volume. Rows correspond to increasing ambient gas density from $30~m_p~\mathrm{cm^{-3}}$ to $3\times10^{3}~m_p~\mathrm{cm^{-3}}$. Columns compare simulations including only DF (left) and those including NDF (right).}
    \label{fig:track}
\end{figure*}

\begin{figure} \centering
\includegraphics[width=1\linewidth]{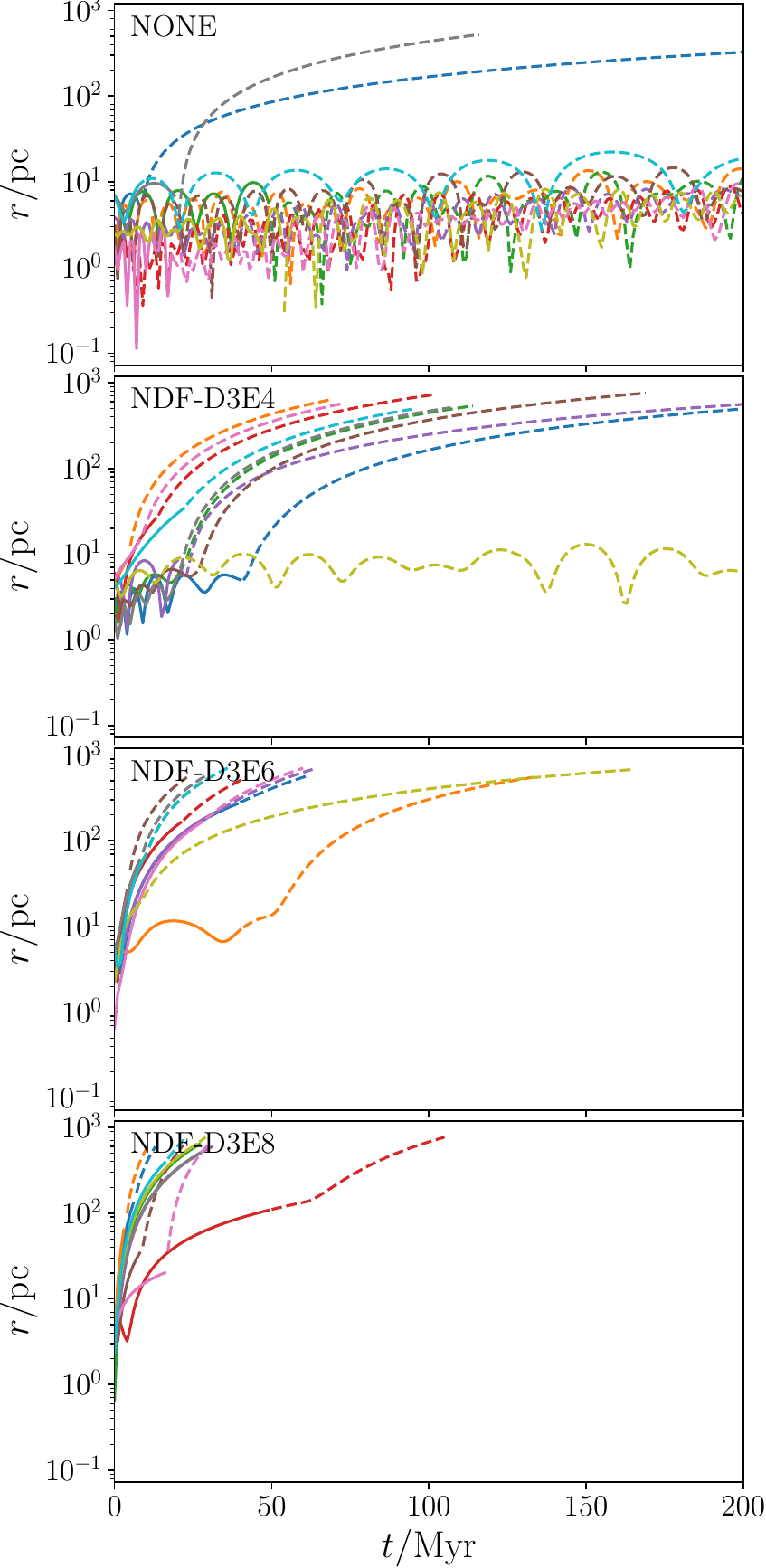}
%    \caption{The track of stars evolving from MS to NS. Results from simulations other than those shown in Figure \ref{fig:track}.}
\caption{Radial trajectories of stars that transition from the MS phase to the NS phase in simulations with other ambient gas densities than those shown in Figure~\ref{fig:track}.
The layout and line styles follow the same conventions as in Figure~\ref{fig:track}. Panels labeled NONE show gas-free reference models, while panels labeled NDF-D3E4, NDF-D3E6, and NDF-D3E8 illustrate the effects of NDF in progressively denser environments.}
    \label{fig:track_others}
\end{figure} 

Figure \ref{fig:track} and \ref{fig:track_others} show the radial trajectories of stars evolve from MS to NS phase, plotted  as a function of time and distance from cluster center. Solid lines denote the trajectories of stars during their MS phase, while dashed lines indicate the post-SN trajectories after the stars have transitioned into NSs. Breaks in the trajectories mark objects that leave the simulation volume.

%Figure \ref{fig:track} compares differences between DF and NDF in the same gas densities. Figure~\ref{fig:track_others} presents results from the remaining simulations that are not included in Figure~\ref{fig:track}. The NDF-affected models demonstrate a striking contrast with DF-only and gas-free scenarios: all NSs experiencing NDF are systematically expelled from the cluster potential, while those in DF and NONE modes predominantly remain bound. This expulsion occurs on timescales significantly shorter than the cluster's natural evaporation time, demonstrating that NDF provides an efficient mechanism for NS removal independent of supernova kicks.

A clear qualitative difference emerges between simulations that include NDF and those that do not. In the DF-only and gas-free (NONE) runs, most NSs remain confined within the inner few to tens of parsecs over the full $200~\mathrm{Myr}$ evolution, with radial excursions comparable to those of their MS progenitors. By contrast, NDF runs show systematic outward migration of NSs after their formation, with a substantially enhanced fraction of objects reaching large radii or exiting the simulated cluster region.

%The efficiency of NS expulsion depends strongly on ambient gas density. At $\rho_0 = 3\times 10^4~ m_p ~\mathrm{cm^{-3}}$, NSs typically require $\gtrsim 100 ~{\rm Myr}$ to escape the cluster, while at $\rho_0 = 3\times 10^6,~3\times 10^8~ m_p~ \mathrm{cm^{-3}}$, this timescale reduces to $\sim 50 ~\rm{Myr}$ and $\sim 25 ~\rm Myr$ respectively. In NDF-D3E8, almost all massive stars escape from the cluster before evolving into compact objects. This density dependence aligns with the theoretical expectation $a_{\mathrm{NDF}} \propto \rho^{1/2}$, confirming that NDF acceleration can overcome gravitational binding energy more rapidly in denser environments.

The efficiency of NS removal via NDF depends strongly on the ambient gas density. At the lowest densities considered (e.g.\ NDF-D3E1 from Figure \ref{fig:track} top right), only a subset of NSs are driven to large radii within $200~\mathrm{Myr}$, while others remain temporarily bound.
At intermediate densities (e.g.\ NDF-D3E2--D3E3 from Figure \ref{fig:track} middle and bottom right), outward migration becomes more pronounced and escape events occur more frequently. At the highest densities explored (e.g.\ NDF-D3E6 and NDF-D3E8 from the third and last panel of Figure \ref{fig:track_others}), nearly all NSs undergo rapid outward displacement, and a significant fraction of massive stars are transported to large radii even before they evolve into compact remnants.

The characteristic timescale for NS displacement shortens systematically with increasing gas density. In the lowest-density NDF runs, NSs typically require $\gtrsim 100~\mathrm{Myr}$ to reach radii comparable to the size of the simulated cluster region, whereas this timescale decreases to $\sim 50~\mathrm{Myr}$ at intermediate densities and to $\lesssim 30~\mathrm{Myr}$ in the highest-density environments. This behavior reflects the increasing effectiveness of NDF in denser gas, which more rapidly overcomes the gravitational confinement of the cluster in our adopted prescription.

Importantly, the systematic outward migration and loss of NSs in the NDF runs occurs even in the absence of natal SN kicks, demonstrating that NDF provides an efficient and independent channel for removing NSs from young, gas-rich star clusters.

\subsection{Distribution of stars in momentum-radius phase space under DF and NDF}

To further elucidate how NDF modifies stellar dynamics at the cluster scale, we examine the distribution of stars in momentum--radius phase space. This representation allows us to distinguish between stars that remain gravitationally confined and those undergoing systematic outward acceleration.

\begin{figure*} \centering
\includegraphics[width=\linewidth]{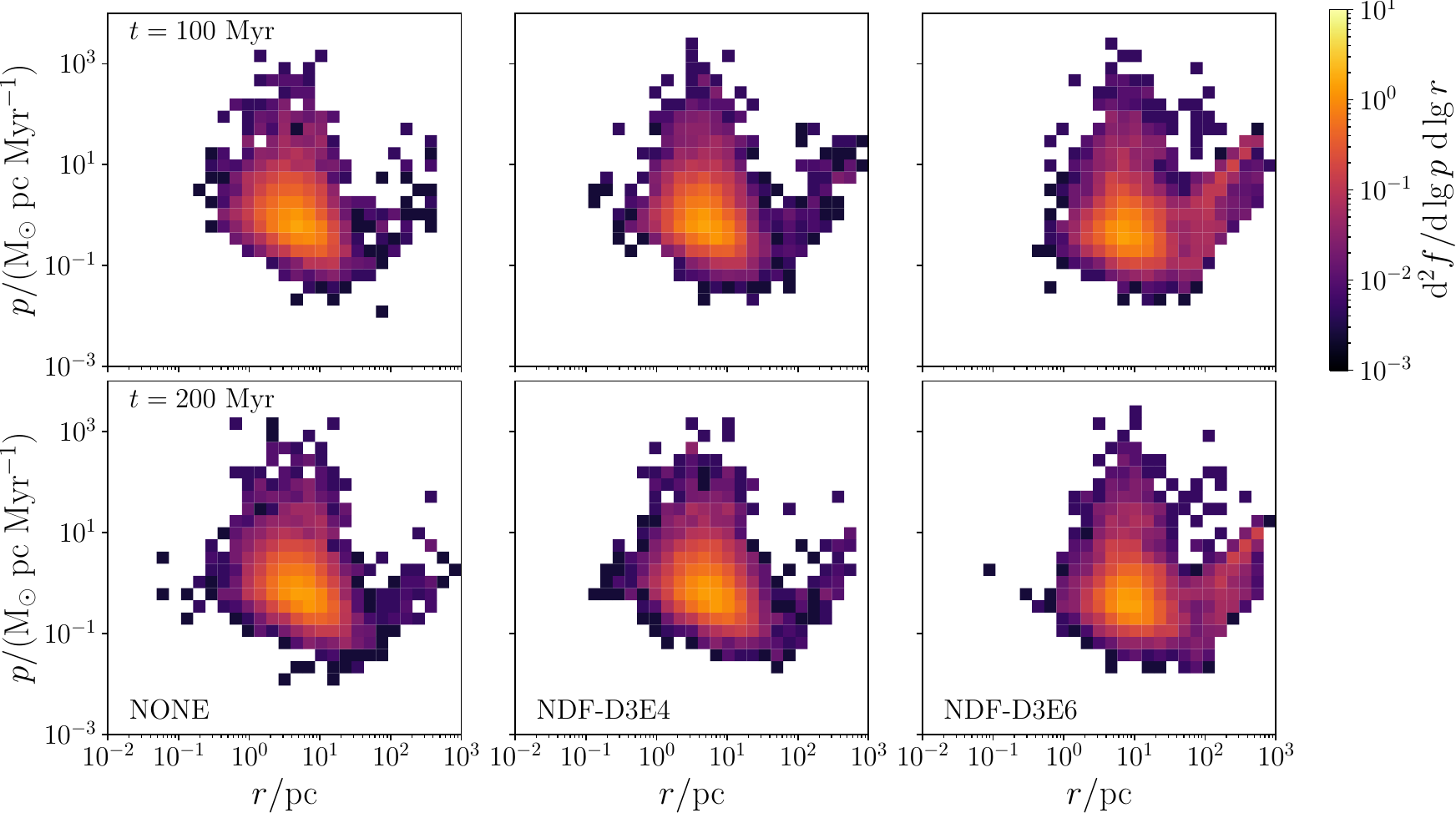}
%    \caption{2D histograms in momentum-radius phase space. The horizontal axis shows the distance of stars from the cluster center of mass, while the vertical axis shows the stars' momentum. The upper and lower rows present snapshots at $t = 100~\mathrm{Myr}$ and $200~\mathrm{Myr}$, respectively. The three columns correspond to NDF-mode results for different gas densities, ranging from $0$ to $3 \times 10^6~m_p~\mathrm{cm^{-3}}$.}
\caption{Two-dimensional histograms in momentum--radius phase space. The horizontal axis shows the distance of stars from the cluster center of mass, while the vertical axis shows the stellar momentum. Color indicates the normalized phase-space density. The upper and lower rows present snapshots at $t = 100~\mathrm{Myr}$ and $200~\mathrm{Myr}$, respectively. The three columns correspond to simulations with increasing ambient gas density, including a gas-free reference case (NONE) and NDF runs with progressively higher densities.}
    \label{fig:p_r}
\end{figure*}

%Figure~\ref{fig:p_r} shows the distribution of stars in momentum-radius phase space. The NDF force creates a new region in momentum-radius phase space at larger distance from the cluster center of mass. The momentum of stars in this region do not exhibit a significant decrease, suggesting that NDF provides a radially outward acceleration that counteracts the gravitational force on stars moving away from the cluster center. As density increases, this new region becomes more prominent in the phase space.

As shown in Figure~\ref{fig:p_r}, simulations including NDF develop a distinct population of stars at large radii that retain substantial radial momentum. This population is largely absent in the gas-free reference case and becomes increasingly prominent as the ambient gas density increases. The presence of stars with relatively high momentum at large distances indicates that NDF supplies a sustained outward acceleration, preventing efficient deceleration by the cluster potential.

The persistence of this high-momentum, large radius population demonstrates that NDF does not merely perturb stellar orbits locally, but instead reshapes the global phase space structure of the cluster. In denser gas environments, the affected region of phase space expands systematically, consistent with the enhanced efficiency of NDF-driven outward transport observed in the stellar trajectory analysis.

\subsection{Radial redistribution of cluster members}

A key question for the long-term dynamics of open clusters is whether gas--star interactions can systematically redistribute cluster members in radius, thereby reshaping the global density profile. To quantify this effect, we examine the radial probability density of stars, $f(r)=\mathrm{d}f/\mathrm{d}r$, at two representative epochs, $t=100~\mathrm{Myr}$ and $200~\mathrm{Myr}$, and compare models spanning different gas densities in the DF and NDF modes.

\begin{figure} \centering
\includegraphics[width=1\linewidth]{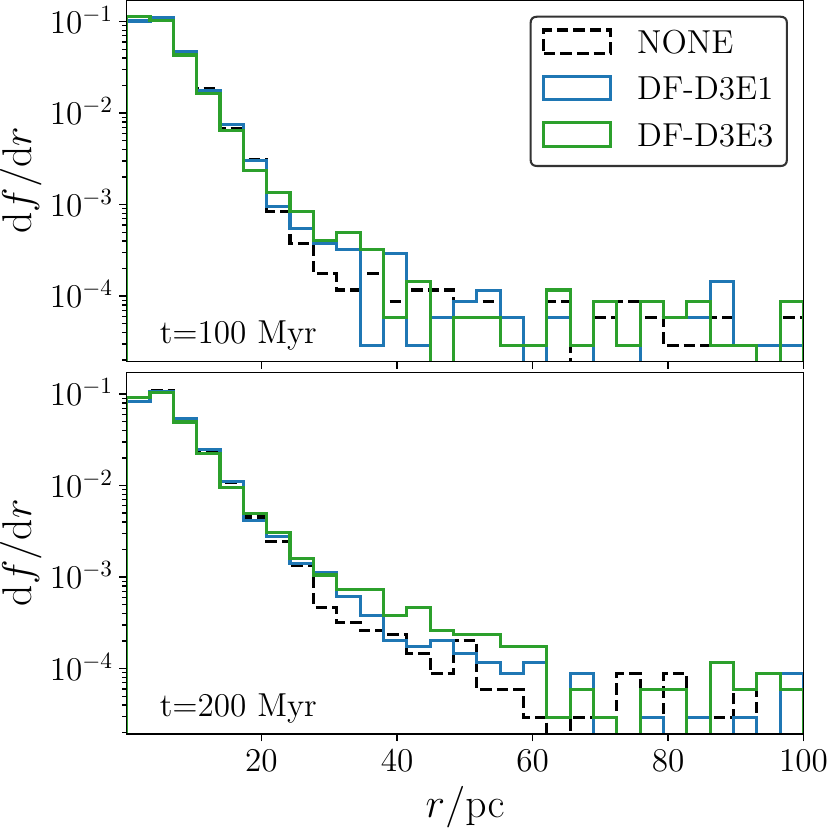}
%    \caption{Radial distribution of stars with different gas density in DF mode. Take the number of stars at distance $[r,r+dr]$ from the OC's center as y-axis. Take the distance of the star from the OC's center $r$ as x-axis. The different sub-figures show the distributions at $t=100~{\rm Myr}$ and $200~\rm Myr$. Each color represents a different gas density.}
\caption{
Radial probability density of cluster members in the DF mode for different gas densities. The quantity shown is $f(r)=\mathrm{d}f/\mathrm{d}r$, where $f(r)$ denotes the fraction of stars located within $[r,r+\mathrm{d}r]$ from the cluster center. Results are shown at $t=100~\mathrm{Myr}$ (upper panel) and $t=200~\mathrm{Myr}$ (lower panel). Different colors correspond to different gas densities, while the dashed curve indicates the no-gas case.
}
    \label{fig:star_distribution_hist_DF}
\end{figure}

\begin{figure} \centering
\includegraphics[width=1\linewidth]{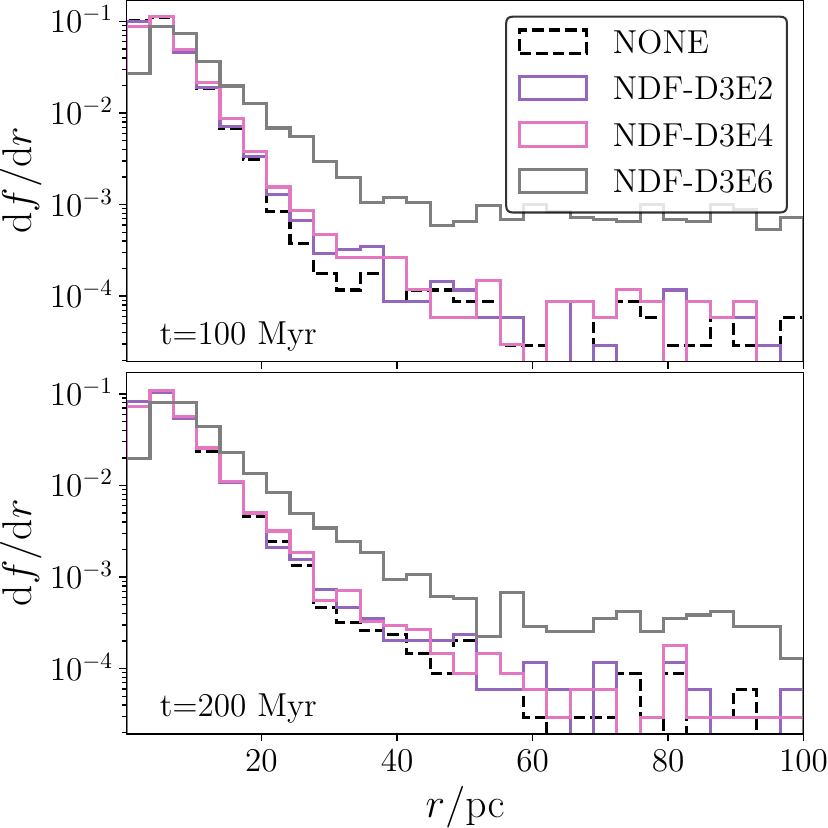}
    \caption{Same as Figure. \ref{fig:star_distribution_hist_DF}, but for the NDF mode.}
    \label{fig:star_distribution_hist_NDF}
\end{figure}

In the absence of gas, as well as in the DF mode, the stellar distribution remains centrally concentrated throughout the evolution, with a pronounced central peak persisting at both epochs. By contrast, clusters evolved in the NDF mode exhibit systematic radial expansion: their radial profiles become progressively shallower, and the peak of $f(r)$ shifts toward larger radii over time.

The strength of this redistribution increases with gas density. In particular, for $\rho = 3\times10^{6}\,m_p\,\mathrm{cm^{-3}}$, the probability density exceeds that of the no-gas case beyond the second radial bin, indicating efficient outward transport of cluster members driven by non-dissipative gas--star interactions. These trends are summarized in Figures~\ref{fig:star_distribution_hist_DF} and \ref{fig:star_distribution_hist_NDF}.

%Figure~\ref{fig:star_distribution_hist_DF} and \ref{fig:star_distribution_hist_NDF} show the radial distributions of stars for different gas densities in the DF and NDF modes, respectively. The impact of gas-star interactions on global cluster structure becomes evident. Initially concentrated profiles evolve differently across interaction modes: while DF and no-gas models maintain centrally peaked distributions, NDF models develop progressively shallower profiles with peaks shifting to larger radii. The probability distribution of $\rho = 3\times 10^6~m_p~{\rm cm^{-3}}$ is significantly higher than that of the no-gas model after the second radial bin.

%This redistribution occurs more rapidly at higher gas densities, with the distribution peak moving to the second radial bin by 100 Myr at $\rho_0 = 3\times 10^4 m_p \mathrm{cm^{-3}}$. At the highest densities, NDF drives a substantial fraction of stars toward the cluster periphery, forming an expanding shell structure that accelerates cluster dissolution.

\subsection{Evolution of cluster scale parameters}

The global structural evolution of OCs can be characterized by the time evolution of their characteristic spatial and kinematic scales. In this section, we quantify how gas--star interactions modify cluster expansion and internal kinematics under different interaction regimes.

\begin{figure*} \centering 
\includegraphics[width=1\linewidth]{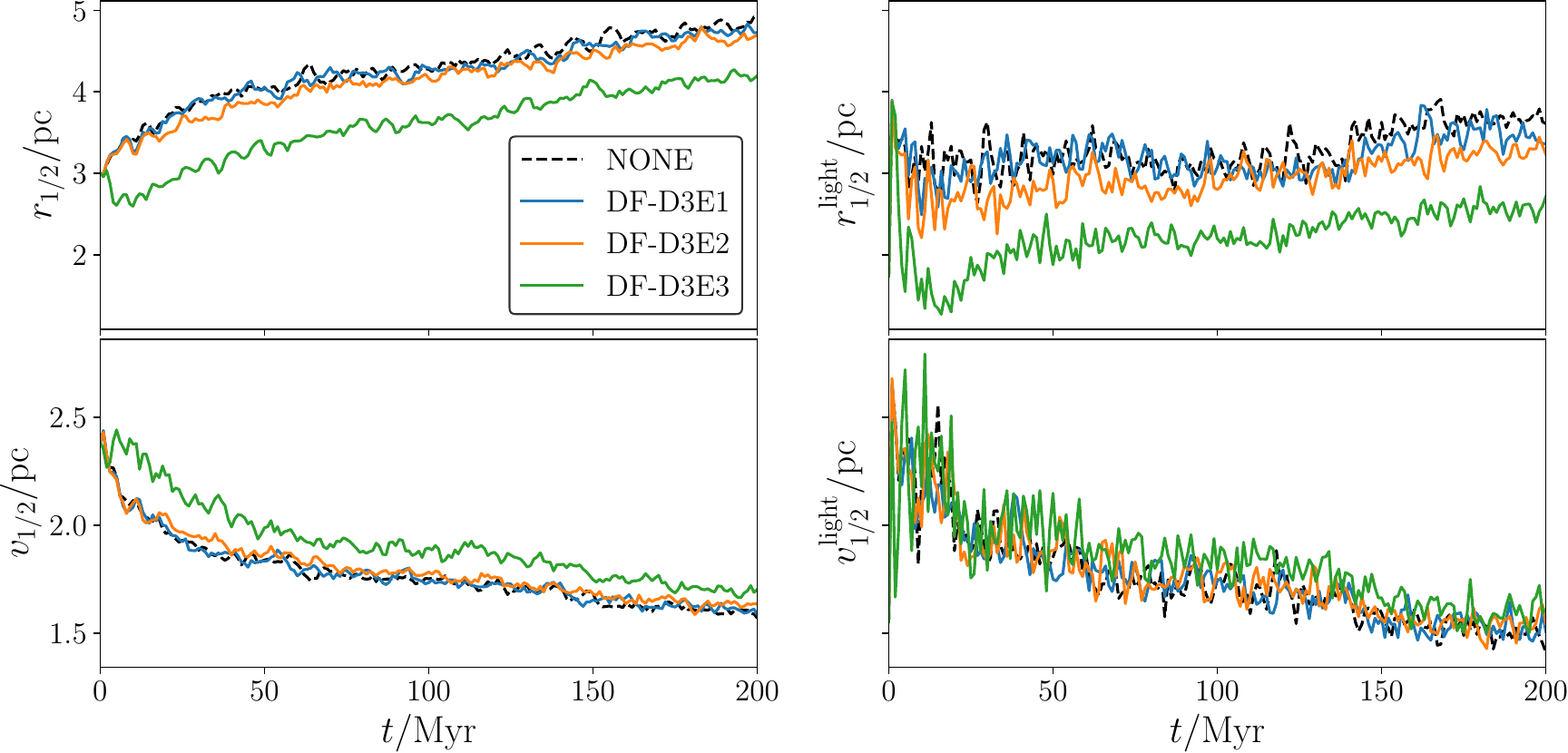}
%    \caption{Half-mass radius, half-light radius, half-mass velocity and half-light velocity as functions of time in DF mode. The black dashed line, blue, orange, and green line correspond to $\rho = 0, 30, 300, 3\times 10^3~m_p~{\rm cm^3}$ respectively.}
\caption{Time evolution of the half-mass radius, half-light radius, and the corresponding velocity scales in the DF mode. Different colors indicate different ambient gas densities, $\rho = 0, 30, 300,$ and $3\times10^{3}\,m_p\,\mathrm{cm^{-3}}$, with the no-gas case shown as a black dashed line.}
    \label{fig:half_DF}
\end{figure*}

\begin{figure*} \centering 
\includegraphics[width=1\linewidth]{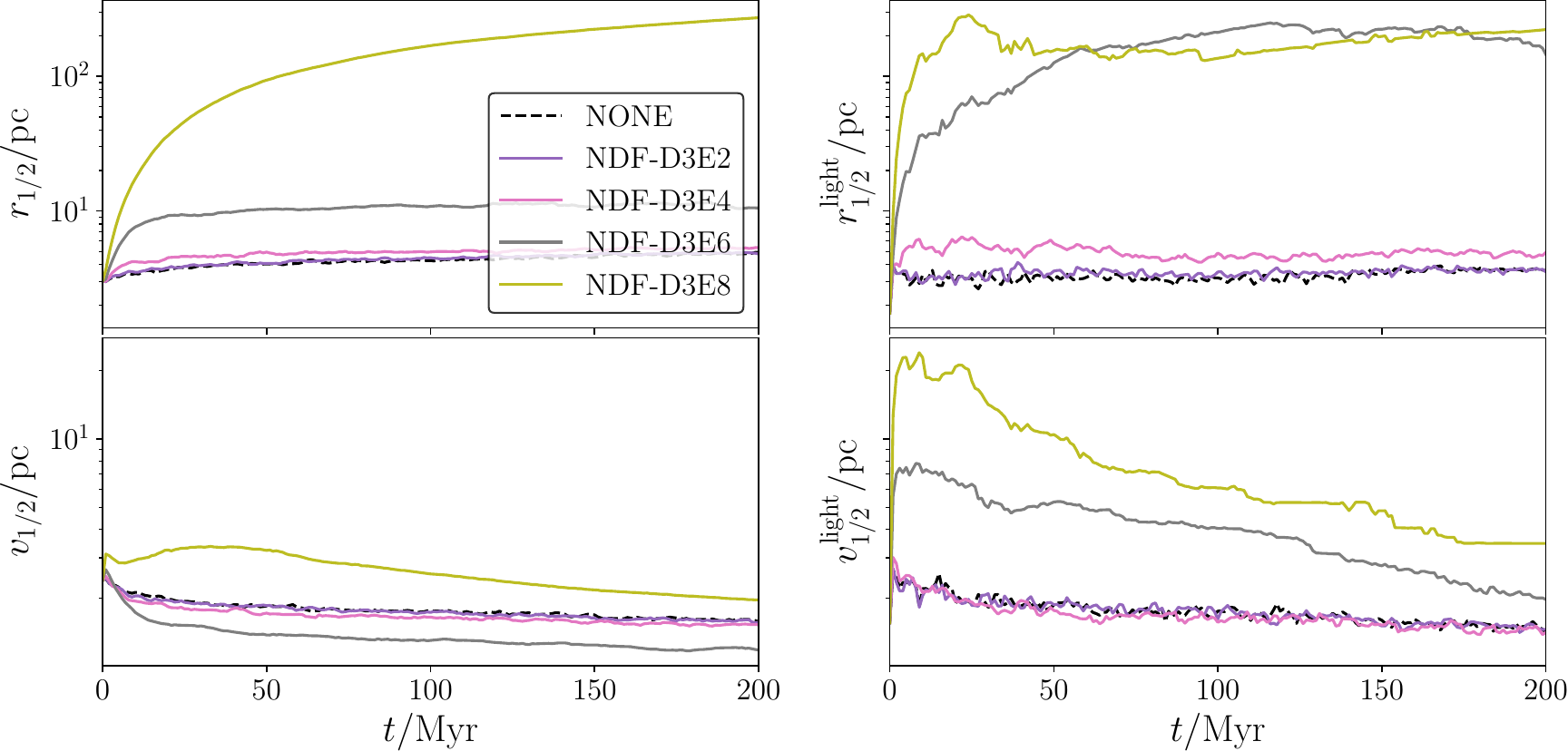}
%    \caption{Similar to Figure \ref{fig:half_DF}, but in NDF mode. The black dashed line, purple, pink, grey, and dark yellow line correspond to $\rho = 0, ~300, ~3\times 10^4, ~3\times 10^6,~3\times 10^8~m_p~{\rm cm^{-3}}$respectively. A logarithmic scale is used to accommodate the large dynamical range.}
\caption{Same as Figure~\ref{fig:half_DF}, but for the NDF mode. A logarithmic scale is used to accommodate the large dynamical range.}
    \label{fig:half_NDF}
\end{figure*}
%两张图的half-mass velocity的趋势并不和half-mass velocity相同，DF的half-light velocity几个密度没有像half-mass velocity那样区分开，NDF half-light velocity和half-mass velocity随密度变化不单调（下面对此进行了简单的解释，不确定是否合适）

% general discription of 2 figure
%The structural evolution of clusters under different interaction regimes is quantified through half-mass and half-light radii $r_{1/2},~r_{1/2}^{\rm light}$. Figure \ref{fig:half_DF} and \ref{fig:half_NDF} show $r_{1/2},~r_{1/2}^{\rm light}$ and $v_{1/2}~,v_{1/2}^{\rm light}$ (half-mass and half-light radii in velocity space). Throughout the simulation, NDF-affected clusters consistently exhibit larger scale radii compared to DF and NONE models across all gas densities. This size difference becomes more pronounced with increasing ambient density, reflecting the enhanced efficiency of NDF in transferring energy to cluster members.
%可直接替换上面一段，Written by Meng for clarity: 
Across all gas densities, clusters evolved in the NDF mode consistently develop larger characteristic radii than their DF and no-gas counterparts (Figures~\ref{fig:half_DF} and \ref{fig:half_NDF}). This systematic size difference becomes more pronounced at higher gas densities, indicating that non-dissipative gas--star interactions act as an effective heating mechanism that transfers kinetic energy to cluster members and promotes global expansion.

% radius in DF mode
%In all cases, half-mass radii increase over time as a result of the OC evaporation. In models with higher gas density, $r_{1/2}$ and $r_{1/2}^{\rm light}$ decreases, while $v_{1/2}$ and $v_{1/2}^{\rm light}$ increases. In intermediate-density environments ($\rho_0 = 3\times 10^3~ m_p ~\mathrm{cm^{-3}}$), DF produces slightly more compact configurations than NONE models, as friction removes orbital energy from stars. 
%可直接替换上面一段，Written by Meng for clarity: 
In the DF mode, the half-mass radius increases gradually with time in all models, reflecting the combined effects of two-body relaxation and cluster evaporation. At fixed evolutionary time, increasing gas density leads to smaller half-mass radii and larger velocity scales, as DF extracts orbital energy from stars and suppresses cluster expansion. In intermediate-density environments ($\rho = 3\times10^{3}\,m_p\,\mathrm{cm^{-3}}$), DF produces slightly more compact configurations than the no-gas case, consistent with the dissipative nature of the interaction.
The half-light radius and half-light velocity exhibit larger fluctuations at early time, likely caused by massive stars with short main-sequence lifetimes.

% difference between half-mass and half-light, DF mode
%The differential evolution of half-mass radius and half-light radius provides additional insights into mass-dependent dynamics. Since half-light radius weights stars by luminosity ($L \propto M^{3.5}$), it responds more sensitively to the spatial distribution of massive stars. In DF models, the retention of massive stars in central regions (due to their stronger friction) maintains relatively small half-light radius values even as half-mass radius increases, creating a measurable disparity between these scale parameters that could serve as an observational diagnostic of DF dominance.

%可直接替换上面一段，Written by Meng for clarity: 
The evolution of the half-light radius differs systematically from that of the half-mass radius, reflecting the mass dependence of cluster dynamics. Because the half-light radius weights stars by luminosity ($L \propto M^{3.5}$), it is more sensitive to the spatial distribution of massive stars. In the DF mode, massive stars experience stronger friction and are preferentially retained in the central regions. As a result, the half-light radius remains relatively small even as the half-mass radius increases, producing a persistent offset between these two scale parameters. This differential behavior provides a potential observational diagnostic of DF-dominated evolution.

% NDF mode
%A significant early-time spike in both the half-mass and half-light radii is observed in the NDF-D3E6 model. This feature is likely driven by massive main-sequence stars that produce strong stellar winds, making the spike in the half-light radius more prominent than that in the half-mass radius. After the short main-sequence lifetimes of these massive stars, they collapse into black holes, and the radii subsequently stabilize.
%For the same reason, the half-mass velocity in the NDF-D3E6 model is larger than in the other three simulations in Figure~\ref{fig:half_NDF} at early times, but it rapidly decreases to the smallest value due to the cluster relaxation. Because the luminosity excludes compact objects, in NDF mode the half-mass velocity with $\rho =3 \times 10^8~m_p~\mathrm{cm^{-3}}$, half-light velocity with $\rho = 3 \times 10^6,~3 \times 10^8~m_p~\mathrm{cm^{-3}}$ remains relatively high, indicating that main-sequence stars gain substantial kinetic energy from the NDF force.

%可直接替换上面一段，Written by Meng for clarity: 
Clusters evolved in the NDF mode exhibit qualitatively different behavior. At early times, a pronounced expansion is observed in both the half-mass and half-light radii for the two highest-density models ($\rho = 3\times10^{6},~3\times10^{8}\,m_p\,\mathrm{cm^{-3}}$). This transient feature is driven by massive MS stars whose strong stellar winds enhance the coupling between the gas and stellar components. Because luminous stars dominate the light budget, the resulting spike is more prominent in the half-light radius than in the half-mass radius.

%可直接替换half-mass velocity那段，Written by Meng for clarity: 
A similar trend is reflected in the velocity scales. At early times, the half-mass velocity in the $\rho = 3\times10^{8}\,m_p\,\mathrm{cm^{-3}}$ NDF model exceeds that of lower-density cases, indicating efficient energy injection into the stellar component. As the most massive stars rapidly evolve and collapse into compact objects, this excess kinetic energy is redistributed through relaxation, leading to a subsequent decline in the velocity scale. Because compact remnants do not contribute to the cluster luminosity, the half-light velocity remains relatively high for $\rho = 3\times10^{6}$ and $3\times10^{8}\,m_p\,\mathrm{cm^{-3}}$, implying that MS stars acquire substantial kinetic energy through non-dissipative gas--star interactions.

%整个Section总结，这一节主要是：
%DF --> dissipative cooling + mass segregation；
%NDF --> effective heating + scale expansion；
%half-light vs half-mass --> 建立动力学与可观测量的connection

\section{Binary dynamics in clusters immersed in gas}
\label{sec:binary}

%Binary systems represent a crucial component of stellar
%populations in open clusters, playing a significant role in
%their dynamical evolution. The presence of ambient gas can
%profoundly alter the orbital properties and survival rates
%of binaries through both dynamical friction (DF) and
%negative dynamical friction (NDF). 

In this section, we
investigate how these gas-star interactions influence the
formation, stability, and evolution of binary systems in
OCs.

To illustrate the physical scales involved, consider an
isolated binary consisting of two MS stars with $m_1=m_2=5~M_{\odot}$ on a circular orbit with separation $r = 5~\mathrm{AU}$. Each star orbits the center of mass with speed
\begin{equation}
v_*=\sqrt{\frac{G m_1}{2r}}\simeq 21.1~\mathrm{km~s^{-1}}.
\end{equation}

%$v_* = \sqrt{GM/2r} \sim 21.1~\mathrm{km~s^{-1}}$. Each star is enveloped by the outflow from its companion, and the local density of the wind at the companion's location is$\rho_0 = \dot{m}/(4\pi r^2 v_{\rm w})$. The resulting NDF acceleration is estimated as$a_{\mathrm{NDF}} = 8.18G\rho_0 R_0 \approx 2.8\times 10^{-14}~\mathrm{cm~s^{-2}}$. The timescale for the binary orbit to double its angular momentum under this acceleration is $\tau \sim v_*/a_{\mathrm{NDF}} \approx 2 \times 10^6~\mathrm{Myr}$, which far exceeds the typical cluster lifetime. This simple estimate suggests that, for typical main-sequence binaries, NDF alone is unlikely to significantly alter the orbit over the cluster’s evolution. However, in denser gas environments or for more powerful outflows, the effect can become substantial.

If the companion outflow sets a local gas density $\rho_0\simeq \dot{m}/(4\pi r^2 v_{\rm w})$ at the location of the accretor, the characteristic NDF acceleration can be estimated as $a_{\rm NDF}\propto G \rho_0 R_0$ (see Section~\ref{sec:method} for the definition of $R_0$). The corresponding timescale to modify the orbital velocity by order unity is
\begin{equation}
\tau_{\rm NDF}\sim \frac{v_*}{a_{\rm NDF}},
\end{equation}
which is typically much longer than the $\sim 2 \times 10^6~\mathrm{Myr}$ lifetime of an open cluster for MS winds and separations of interest. This estimate suggests that, for typical MS binaries, NDF acting in isolation is unlikely to reshape the orbit over the cluster lifetime. In denser gas environments, or for systems embedded in unusually strong outflows, the effect can become non-negligible.

To ensure computational efficiency and avoid the omission of wide pairs, we identify candidate binaries with a maximum separation $s_{\max}=5~\mathrm{pc}$ (Section~\ref{sec:method-bin}), followed by our
standard boundness criteria. As the separation of widest binary in the cluster is $\sim 10^4~{\rm AU}\sim 0.05~{\rm pc}$ \citep{2003gmbp.book.....H, 2020ApJ...904..113R}, we have also tested $s_{\max}=0.1~\mathrm{pc}$ and obtained very similar results, indicating that our conclusions are not sensitive to this choice.

\begin{figure*}
\centering
\includegraphics[width=1\linewidth]{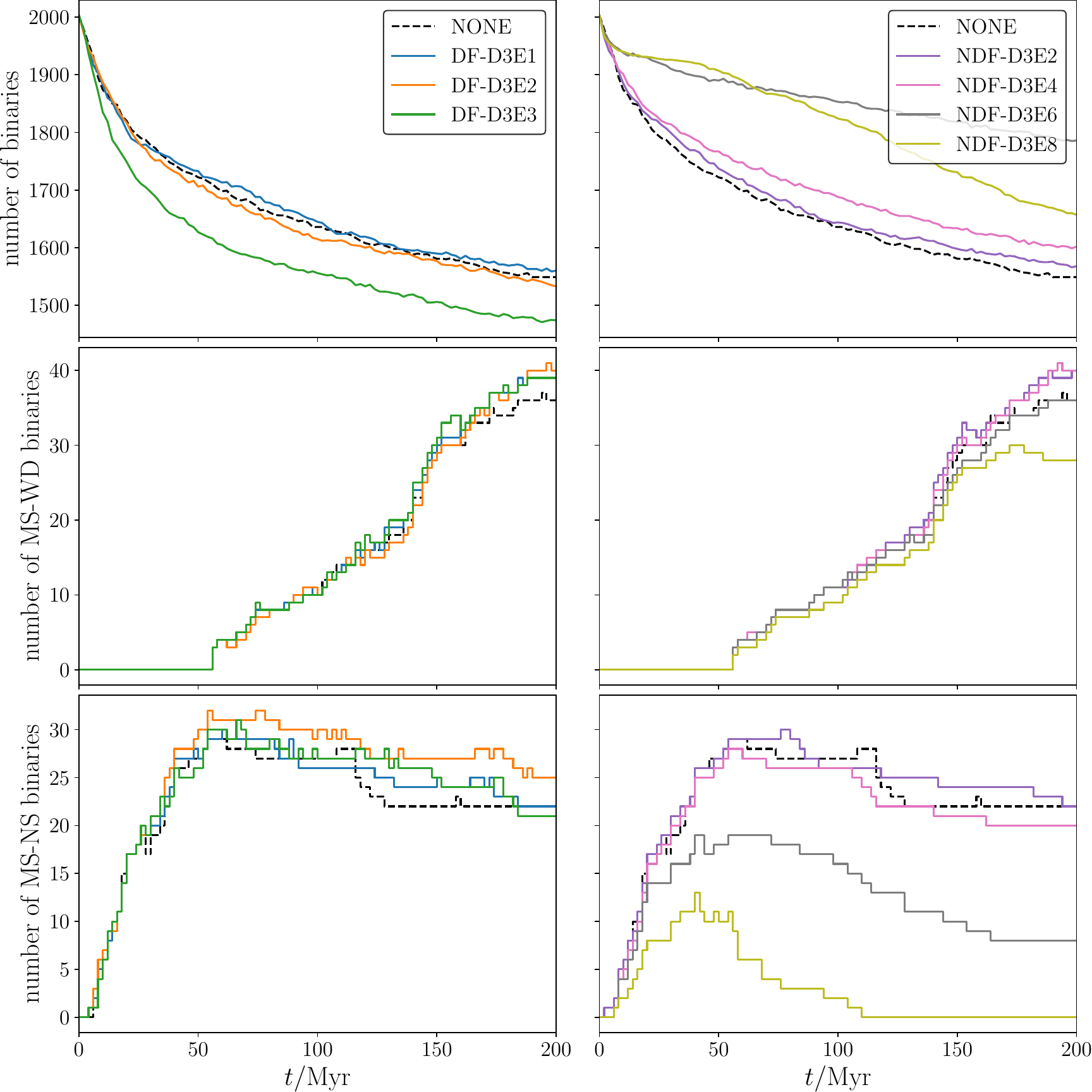}
\caption{ Temporal evolution of the total number of binaries, as well as the numbers of MS--WD and MS--NS binaries 
in the cluster for different gas densities and interaction models. 
The left column shows results in DF mode, while the right 
column shows results in NDF mode.
At high gas densities, DF suppresses survival of binaries, while NDF
maintains a larger number of surviving binaries. }
\label{fig:binary_number}
\end{figure*}

The time evolution of the binary population is summarized in Figure~\ref{fig:binary_number}. At moderate densities ($\rho = 30$--$300~m_p~\mathrm{cm^{-3}}$), both DF and NDF have negligible effects on the total number of bound binaries. At higher gas densities, the binary population exhibits a clear mode dependence: DF significantly reduces the number of surviving binaries, whereas NDF increases it. Moreover, the enhancement in the NDF mode becomes increasingly pronounced with increasing gas density.

% explain why DF decrease the number of binary and NDF increase;
We classify binaries by their binding energy, $E_{\rm b} = -E_{\rm tot}>0$, where $E_{\rm tot}$ is defined in Eq.~\eqref{eq:E_tot}. Hard binaries are defined as systems whose binding energy exceeds the local average kinetic energy of stars in the cluster ($E_{\rm b}>\overline{E}_{\rm k}$), whereas soft binaries have binding energies below the local average kinetic energy ($0<E_{\rm b}<\overline{E}_{\rm k}$) and are therefore easily disrupted by encounters. Heggie’s law \citep{1975MNRAS.173..729H} states that hard binaries tend to become harder and soft binaries tend to become softer as a result of multi-body encounters. DF from the ambient gas is expected to increases the stellar density and encounter frequency, which likely enhances this effect. In contrast, NDF leads to a more diffuse stellar distribution, which may reduce the encounter rate and is consistent with an increased survival probability of soft binaries. 

To apply this classification in an inhomogeneous cluster, we estimate the local kinetic energy scale around each binary using the stellar neighborhood of that binary. Specifically, for each binary we select the 100 stars nearest to the binary center of mass and compute the mean kinetic energy of these neighbors
\begin{equation}
\begin{aligned}
    \overline{E}_{\rm k}&=\frac{1}{N}\sum_{i=1}^{N}\frac12 m_i |\Delta v_i|^2,\\
    \Delta v_i &=  v_i - \dfrac{\sum_N m_iv_i}{\sum_N m_i} , 
\end{aligned}
\end{equation}
where $\Delta v_i$ is measured relative to the local center-of-mass
velocity of the stellar neighborhood surrounding the binary and $N=100$. The results are insensitive to choosing $N = 50\sim200$. This choice provides a practical, spatially local proxy for the typical kinetic energy scale relevant for encounters involving by that binary, and avoids relying on a single cluster-averaged velocity scale, which can be misleading once the system develops strong radial gradients in density and velocity dispersion.

\begin{figure*}
    \centering
    \includegraphics[width=1\linewidth]{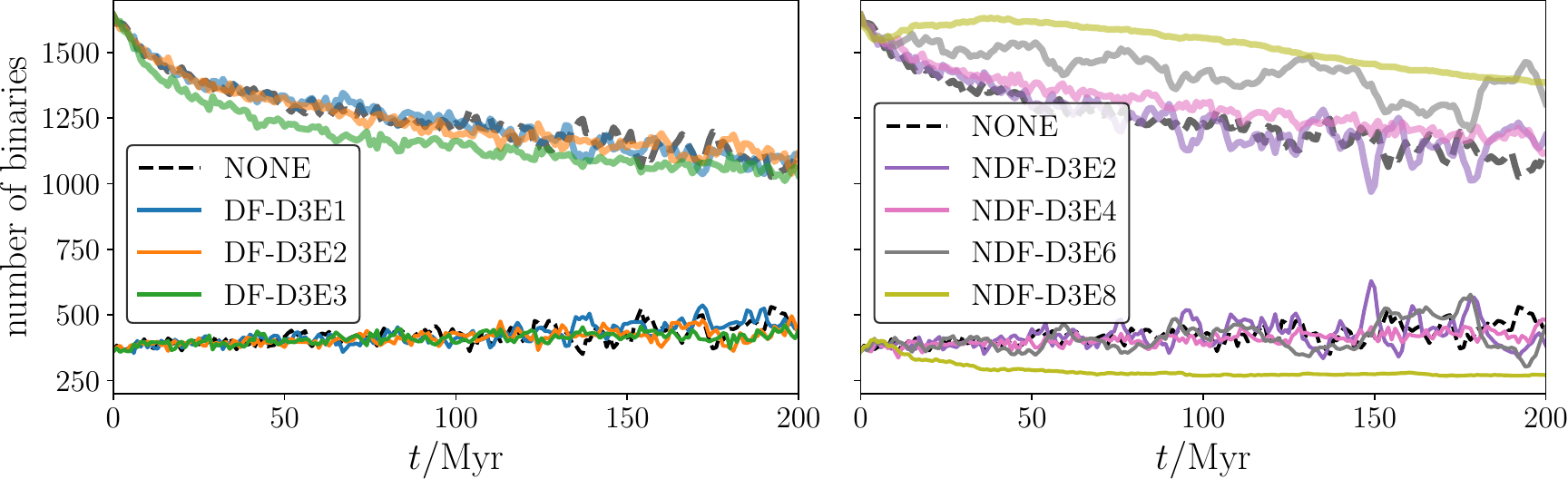}
    \caption{
    %Temporal evolution of the number of soft binaries (thick lines) and hard binaries (thin lines). The left panel shows results in DF mode while the right panel shows results in NDF mode.
    Temporal evolution of the numbers of dynamically soft (thick lines) and hard (thin lines) binaries. Colors denote different ambient gas densities. The left and right panels correspond to models including DF and NDF, respectively.
    }
    \label{fig:soft_hard}
\end{figure*}

Figure \ref{fig:soft_hard} shows the temporal evolution of the numbers of soft and hard binaries in our simulations.
In DF mode, the number of soft binaries decreases steadily with time. In the DF-D3E3 model, the soft-binary population exhibits a significantly smaller number of soft binaries than both the gas-free case and DF models at lower gas densities. In NDF mode, the number of soft binaries increases with increasing gas density. In the highest-density NDF model, non-dissipative gas--star interactions substantially elevate the local kinetic energy scale, shifting the hard-soft boundary to higher binding energies. As a result, many binaries are reclassified as dynamically soft,
%some hard binaries to transition into soft binaries, 
resulting in a reduced hard-binary population and an enhanced soft-binary population.

We further classify binaries by their constituent stellar types, including MS--MS, MS--NS, MS--WD, and MS--BH systems (numbers of MS--WD and MS--NS binaries are shown in Figure~\ref{fig:binary_number}). At low to intermediate gas densities, the number of MS--NS binaries in the DF and NDF model show no significant difference with gas-free model. When $\rho=3\times 10^6 ~m_p~{\rm cm^{-3}}$, the number of MS--NS binaries is much lower than in the DF or gas-free models, whereas MS--WD binaries are almost unaffected by star-gas interaction. 
%This is because NSs experience strong NDF due to their powerful winds, leading to orbital expansion and disruption, while WDs-lacking outflows in our model-are less affected. 
In our simulations, NSs are assigned a strong NDF contribution as an effective representation of momentum and energy injection by pulsar winds, which leads to efficient orbital expansion and disruption of MS--NS binaries. In contrast, WDs, which do not produce comparable outflows in our model, are less
affected. When gas density goes up to $3\times 10^8~m_p~{\rm cm^{-3}}$, the number of MS--NS binaries become even smaller. The effect of winds in MS phase becomes significant at the highest gas density, resulting in a smaller number of MS--WD binaries than in other NDF models at $t\to 200~\rm Myr$. For the same reason, the number of binaries in NDF-D3E8 becomes smaller than in NDF-D3E6 after $\sim 100~\mathrm{Myr}$.

Because the MS lifetimes of low mass stars in our simulations (e.g. $\lesssim 1~M_{\odot}$) are much longer than the total simulation time of $200~\rm Myr$, the number of MS--WD binaries continues to grow after $\sim 50~\rm Myr$, when the most massive WD progenitors ($\sim 8~M_{\odot}$) evolve off the MS. In contrast, once the lowest-mass NS progenitors complete their MS evolution, the number of MS--NS binaries ceases to increases. 
% At very high densities ($\rho \gtrsim 3\times10^4~m_p~\mathrm{cm^{-3}}$), binaries involving compact objects become rare, as the cluster evaporates rapidly and stars do not have sufficient time to form or maintain bound systems.
MS--NS binaries are preferentially removed over time by NDF-driven orbital expansion, while MS--WD binaries are more likely to survive and accumulate. This behavior implies that NDF can significantly alter the relative populations of different binary types in OCs, potentially leading to an enhanced abundance of WD binaries relative to NS binaries.

\begin{figure*}
\centering
\includegraphics[width=\linewidth]{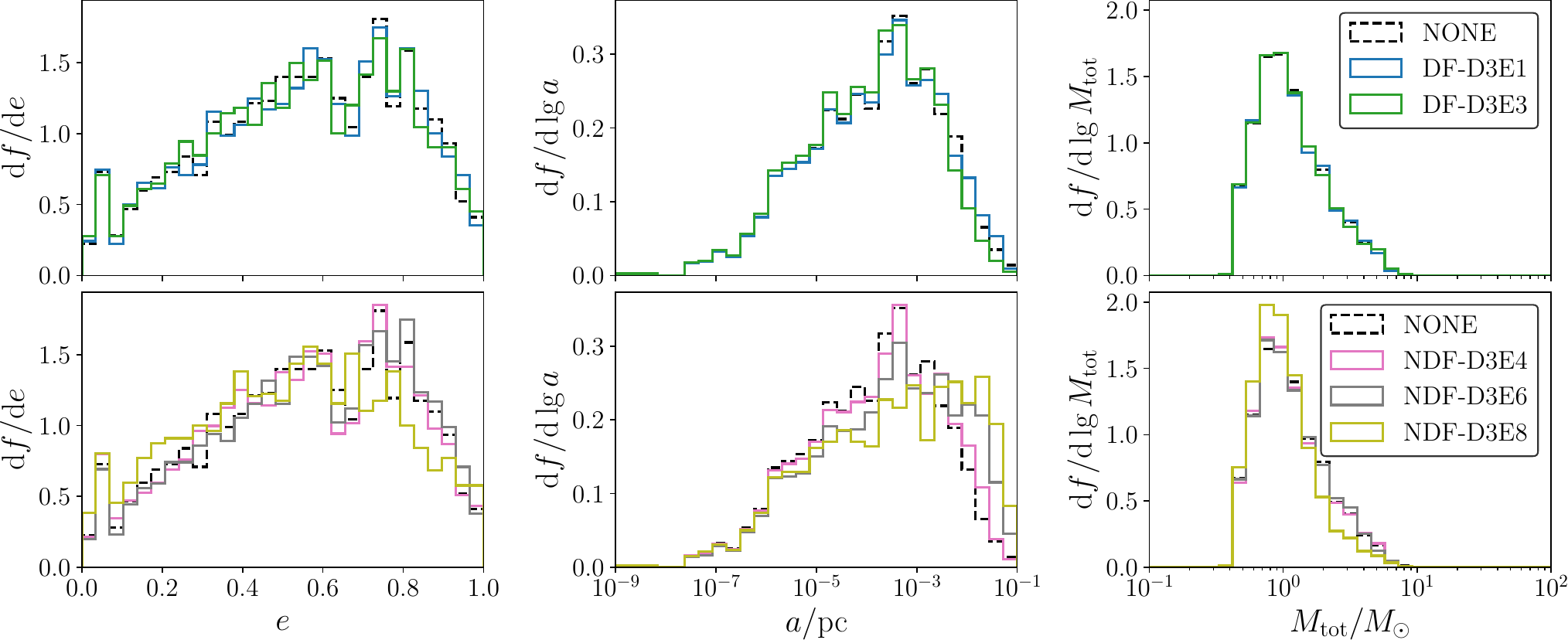}
\caption{ Distributions of orbital parameters for binaries 
  at $t = 200~\mathrm{Myr}$ in different gas-star interaction 
  models and ambient gas densities. 
  The parameters include eccentricity $e$ (left column),
  semi-major axis $a$ (middle column) and total mass of binary system $M_{\rm tot}$ (right column).
  }
\label{fig:parameter}
\end{figure*}

We also analyze the orbital parameters of the surviving
binaries, including eccentricity $e$, semi-major axis $a$, total
mass $M_{\rm tot}$ and orbital inclination $i$.
The distribution of $\cos{i}$ is essentially uniform, which indicates that the cluster remains nearly spherically symmetric throughout its evolution.
$e,~a$ and $M_{\rm tot}$ are shown in Figure~\ref{fig:parameter}. Results of NDF-D3E6 and NDF-D3E8 show that NDF tends to preserve
binaries with larger semi-major axes, producing wider systems, decreasing the binding energy of binary systems. 
This is demonstrated in the study of a separate binary system \citep{2022ApJ...932..108W} that outflows in binary systems exert positive torques and expand the orbits.
Binaries in NDF-D3E8 also exhibit smaller $e$ and smaller $M_{\rm tot}$ than those in other models.  %Binaries containing massive stars are more easily disrupted by NDF force due to their stronger stellar wind. %This leads to more circular orbits and smaller stellar radii within the surviving binaries, thereby preventing stellar collisions and mergers.
This reflects a preferential disruption of binaries containing more massive stars, which experience stronger non-dissipative interactions in our model. As a result, the surviving binary population is biased toward more circular orbits and lower total masses, reducing the likelihood of stellar collisions and mergers.

These results highlight the dual role of gas in binary
evolution: DF tends to suppress the survival of binaries by
enhancing stellar encounters, whereas NDF increases the survival probability of wide and soft
binaries by dynamically heating the cluster and reducing encounter rates. The net
outcome depends sensitively on the ambient gas density and on the properties of stellar outflow adopted in the model. In sufficiently dense environments, gas--star interactions may dominate the dynamical evolution of binaries, with potential observational implications for the binary fraction, orbital period distribution, and compact object merger rates in stellar clusters.

% \begin{sidewaysfigure*}
% \centering
% \includegraphics[width=8in,keepaspectratio]{\figdir/binary_class.pdf}
% \caption{ Number of binaries of different types as a
% function of time. From top to bottom: MS-MS, MS-NS, MS-WD,
% and MS-BH binaries. Blue, yellow, and green curves
% correspond to the NDF, DF, and gas-free models,
% respectively. NDF reduces the number of MS-NS binaries,
% while MS-WD binaries are less affected.  }
% \label{fig:class_binary}
% \end{sidewaysfigure*}

\section{Discussion and summary}
\label{sec:summary}

This study employs the open-source \texttt{PeTar} code to
simulate the dynamical evolution of a massive open cluster
initially comprising 10,000 stars, and systematically
investigating the effects of different gas-star interaction
regimes across a wide range of ambient gas densities. We track both the global structural evolution of the cluster and the
detailed behavior of binary populations, with particular
attention to systems containing MS stars paired
with compact objects (WDs, NSs, and BHs).

The simulations reveal fundamentally contrasting roles of
DF and NDF in shaping cluster structure and binary populations. DF drives mass segregation by extracting orbital energy from stars, causing
systematic orbital decay and enhanced stellar densities in
the cluster core. In contrast, NDF continuously injects kinetic energy
into stellar orbits, accelerating cluster expansion and
promoting evaporation. Stars subject to NDF migrate outward,
producing extended stellar distributions and progressive
depletion of the cluster core.
% This concentration facilitates binary
% formation through dissipative encounters, with the binary
% fraction increasing monotonically with ambient gas
% density. At the highest densities explored
% ($\rho = 3\times10^{8}~m_p~\mathrm{cm^{-3}}$), DF produces a
% central binary-rich region where the number of bound pairs
% temporarily exceeds the stellar count, indicating rapid
% gas-assisted capture.

%Conversely, NDF accelerates cluster evaporation by continuously injecting energy into stellar orbits. Stars experiencing NDF migrate outward, forming an extended outer envelope that progressively depletes the cluster core. This systematic energy injection forms wider and more circular binary systems, as the additional orbital energy overcomes gravitational binding. Binary systems containing NSs exhibit the most pronounced sensitivity to NDF, with MS-NS binaries declining to near-zero abundance within 100 Myr at $\rho =3\times10^{8}~m_p~\mathrm{cm^{-3}}$ due to the powerful winds and consequent strong NDF acceleration of NS components.

At the binary level, NDF preferentially promotes wider and less tightly bound systems by increasing orbital energy, while also tending to circularize orbits. Binary systems containing NSs exhibit the strongest response to NDF. In our simulations, MS--NS binaries decline to nearly zero abundance within $\sim100$ Myr at $\rho = 3\times10^{8}\,m_p\,\mathrm{cm^{-3}}$, reflecting the strong effective NDF acceleration assigned to NSs as a representation of momentum and energy injection by pulsar winds.

The systematic removal of NSs via NDF provides a continuous depletion channel that operates in addition to natal SN kicks. This mechanism may contribute to the observed paucity of NSs in OCs and to their spatial distribution throughout galactic disks. 
Additionally, the density-dependent acceleration of cluster dissolution suggests that encounters with dense molecular clouds could significantly shorten cluster lifetimes, potentially explaining the rapid disruption observed in some young clusters. The distinctive structural evolution under NDF versus DF, particularly the differential behavior of half-mass versus half-light radii, provides observable signatures that could help identify dominant gas-star interaction mechanisms in real clusters.

The interstellar medium densities explored span from typical
Milky Way conditions ($30~m_p~\mathrm{cm^{-3}}$) to extreme
environments such as AGN disks
($3\times10^{8}~m_p~\mathrm{cm^{-3}}$). While the latter may
not represent typical OC habitats, these
high-density simulations serve as controlled scaling experiments that clarify the dependence of gas-star interaction effects on ambient density. The low- to intermediate-density regimes ($30$-$300~m_p~\mathrm{cm^{-3}}$), which are more relevant for typical OCs, already demonstrate that gas interactions can significantly influence binary evolution timescales and population demographics.

% compare this work with Muxin Liu's
Our results are broadly consistent with cluster-scale
dynamical evolution reported by \citet{2025ApJ...978...87L}. An important improvement of the present study is the more robust numerical treatment of the binaries. 
%The relatively low number of surviving binaries reported in \citet{2025ApJ...978...87L} can be attributed mainly to two factors that are mitigated in our simulation: the limited cluster size, which enhances stochastic effects, and the limitations of the \texttt{REBOUND} code in handling close encounters. In such cases, strong gravitational force within binaries may lead to artificial disruption due to integration errors, as illustrated by the rapid dissolution of most primordial binaries at the beginning of the S1500b simulation. Meng: 和别人工作比较的时候，语气要缓和哈，我重写了这句话。另外我想还是不要捎带REBOUND code，好像code不好一样哈哈，其实还是很solid的code
The relatively low number of surviving binaries in \citet{2025ApJ...978...87L} likely reflects the combined effects of smaller particle numbers, which amplify stochasticity, and the use of integration schemes not optimized for long-term resolution of close encounters. In contrast, \texttt{PeTar} is specifically designed for cluster-scale simulations, with accurate handling of binaries, triples, and higher-order systems, leading to improved overall numerical reliability. Consistent with \citet{2025ApJ...978...87L}, we find no evidence that NDF universally suppresses the binary population, nor that DF leads to a substantial binary enhancement over the simulation timescales. Our simulations extend these conclusions by resolving how gas-star interactions reshape binary orbital parameters and subtype demographics.

% In summary, gas-star interactions fundamentally reshape
% binary populations in open clusters through two competing
% mechanisms: DF enhances binary formation and hardening in
% cluster cores, while NDF disrupts binaries, particularly
% those containing stars with powerful winds. These effects
% exhibit strong density dependence and create observable
% signatures in binary demographics, including the
% characteristic scarcity of MS-NS binaries. Future work
% incorporating realistic outflow geometries,
% temperature-dependent gas physics, and comparison with
% larger observational samples will further refine our
% understanding of how gaseous environments influence the life
% cycle of binary systems in stellar clusters.

Several caveats accompany our interpretation. The adopted cluster size ($N=10,000$)
exceeds typical OC populations, and approaches the lower end of globular clusters populations, potentially
amplifying dynamical effects and binary statistics. The
assumption of constant gas temperature ($T=100$ K) neglects
realistic temperature-density correlations and heating from
stellar radiation fields, which could modify the effective
sound speed and Mach number dependencies in Eq. \eqref{eq:a_DF}
and \eqref{eq:a_NDF}. Stellar winds, SN explosions, and the gravitational field of stars can also influence the gas density distribution. 
%Stellar evolution prescriptions, particularly for post-MS phases, incorporate significant uncertainties in mass-loss rates and wind properties that propagate into the calculated NDF accelerations.
Uncertainties in stellar evolution prescriptions, particularly wind properties and post-main-sequence mass loss, also propagate directly into the estimated NDF strength.

%The outflow geometry represents another critical simplification. Our model assumes isotropic winds, whereas real systems often exhibit collimated outflows or jets. \citet{Li_NDF_2020} demonstrated that anisotropic outflows can suppress or eliminate the NDF effect, as directional ejection fails to establish the symmetric underdense cavity required for net acceleration. In such scenarios, the friction force remains positive (decelerating) though reduced in magnitude compared to the no-outflow DF case. Consequently, the NDF effects reported here may represent upper limits, with realistic anisotropic outflows potentially diminishing the significance of NDF in actual cluster environments. The impact of beamed outflows and jets warrants dedicated investigation in future work.

Finally, our treatment assumes isotropic outflows, whereas real stellar winds and jets are often anisotropic. \citet{Li_NDF_2020} showed that strongly collimated outflows can suppress or eliminate the NDF effect by preventing the formation of a symmetric underdense cavity. In such cases, the net force remains decelerating, though reduced relative to the standard DF case. The NDF effects reported here should therefore be regarded as upper limits, and the impact of anisotropic outflows warrants dedicated future study.

Although direct observational constraints on the survival boundary of wide and dynamically soft binaries in open clusters remain limited, several lines of evidence suggest that wide systems can exist under specific environmental conditions. High-angular-resolution imaging surveys have revealed the presence of massive companions at separations up to $\sim 10^{3}$--$10^{4}\,\mathrm{AU}$ in young associations such as Cyg~OB2 \citep{2014AJ....147...40C}. More generally, multiplicity surveys of nearby star-forming regions show substantial populations of wide systems over separations of $\sim 3$--$5000\,\mathrm{AU}$, providing an empirical baseline for the initial wide-binary inventory prior to significant dynamical processing \citep[e.g.,][]{Kraus2011}. These observations suggest that environmental effects beyond simple stellar-encounter rates may play a role in regulating the survival of soft binaries \citep{2007ApJ...662..413K}.

%Observational evidence provides encouraging validation of
%our findings. \citet{gaia_wd} report approximately 3,200
%candidate MS-WD binaries in Gaia DR3, with 110 confirmed
%through ultraviolet excess, compared to only $\sim$50 MS-NS
%candidates with 21 confirmations \citep{gaia_ns}. This
%order-of-magnitude disparity aligns with our NDF model
%predictions, wherein MS-NS binaries are efficiently
%disrupted while MS-WD systems persist. Furthermore, our
%results offer a natural explanation for the difficulty in
%forming MS-NS binaries in low-mass clusters noted by
%\citet{2024OJAp....7E..39T}. In NDF-affected models, MS-NS
%binaries rapidly disappear while MS-BH binaries maintain
%stable populations, creating the observed deficit of MS-NS
%systems relative to theoretical expectations that neglect
%wind-mediated dynamics.

%\begin{acknowledgments}
\section*{Acknowledgments}

The computational resources supporting this work are provided by the Kavli Institute for Astronomy and Astrophysics (KIAA).

Special thanks are extended to Chao Lin for his contributions to the development of the dynamical friction module in \texttt{PeTar}.  We thank our colleagues Huan Yang and Chuizheng Kong for helpful discussions, and we are also grateful to Sanaea Rose for insightful discussions on the dynamical definitions and classification of binary systems.

L.W. acknowledges support from the National Natural Science Foundation of China (Grant Nos. 12233013 and 12573041), the High-level Youth Talent Project (Provincial Financial Allocation; Grant No. 2023HYSPT0706), and the Fundamental Research Funds for the Central Universities at Sun Yat-sen University (Grant No. 2025QNPY04).

M.S. acknowledges support from the National Natural Science Foundation of China (NSFC) through the Fundamental Science Center for Nearby Galaxies (Grant No. 12588202), a project based on LAMOST and FAST, as well as from an additional NSFC grant (Grant No. 12503041; PI: Meng Sun). M.S. also acknowledges support from the Gordon and Betty Moore Foundation through Grant GBMF8477 (PI: Vicky Kalogera).

R.S. acknowledges support from the German Research Foundation (DFG; Grant No. Sp 345/24-1) and from the NSFC (Grant No. 12473017). The International Cooperation Office of the National Astronomical Observatories of the Chinese Academy of Sciences (NAOC, Beijing) supported the work of R.S. in China during 2023–2025. Since 2026, support from the Chinese Academy of Sciences President’s International Fellowship Initiative for Visiting Scientists (PIFI; Grant No. 2026PVA0089) is acknowledged.
%\end{acknowledgments}

\vspace{5mm}

\software{ \texttt{PeTar} \citep{2020MNRAS.497..536W}, \texttt{McLuster} \citep{mcluster}, \texttt{numpy} \citep{2020Natur.585..357H}, \texttt{Matplotlib} \citep{2007CSE.....9...90H}.}

\bibliography{oc_bin}
\bibliographystyle{aasjournal}
\end{CJK*}
\end{document}